\documentclass[suppldata]{interact}

\usepackage{epstopdf}
\usepackage{subfigure}

\usepackage{natbib}
\bibpunct[, ]{(}{)}{;}{a}{}{,}
\renewcommand\bibfont{\fontsize{10}{12}\selectfont}

\theoremstyle{plain}

\theoremstyle{definition}

\theoremstyle{remark}

\usepackage{eurosym}
\usepackage{hyperref}
\usepackage{booktabs,arydshln}
\makeatletter 
\def\adl@drawiv#1#2#3{%
	\hskip.5\tabcolsep
	\xleaders#3{#2.5\@tempdimb #1{1}#2.5\@tempdimb}%
	#2\z@ plus1fil minus1fil\relax
	\hskip.5\tabcolsep}
\newcommand{\cdashlinelr}[1]{%
	\noalign{\vskip\aboverulesep
		\global\let\@dashdrawstore\adl@draw
		\global\let\adl@draw\adl@drawiv}
	\cdashline{#1}
	\noalign{\global\let\adl@draw\@dashdrawstore
		\vskip\belowrulesep}}
\makeatother

\begin{document}


\title{An adaptive route choice model for integrated fixed and flexible transit systems}

\author{
\name{David Leffler\textsuperscript{a}\thanks{CONTACT D. Leffler Email: dleffler@kth.se}, Wilco Burghout\textsuperscript{a}, Oded Cats\textsuperscript{a,b} and Erik Jenelius\textsuperscript{a}}
\affil{\textsuperscript{a}Division of Transport Planning, KTH Royal Institute of Technology, Stockholm, Sweden; \textsuperscript{b}Department of Transport and Planning, Delft University of Technology, Delft, The Netherlands}
}

\maketitle

\begin{abstract}
Over the past decade, there has been a surge of interest in the transport community in the application of agent-based simulation models to evaluate flexible transit solutions characterized by different degrees of short-term flexibility in routing and scheduling. A central modeling decision in the development of an agent-based simulation model for the evaluation of flexible transit is how one chooses to represent the mode- and route-choices of travelers. The real-time adaptive behavior of travelers is intuitively important to model in the presence of a flexible transit service, where the routing and scheduling of vehicles is highly dependent on supply-demand dynamics at a closer to real-time temporal resolution. We propose a utility-based transit route-choice model with representation of within-day adaptive travel behavior and between-day learning where station-based fixed-transit, flexible-transit, and active-mode alternatives may be dynamically combined in a single path. To enable experimentation, this route-choice model is implemented within an agent-based dynamic public transit simulation framework. Model properties are first explored in a choice between fixed- and flexible-transit modes for a toy network. The framework is then applied to illustrate level-of-service trade-offs and analyze traveler mode choices within a mixed fixed- and flexible transit system in a case study based on a real-life branched transit service in Stockholm, Sweden.

\end{abstract}


\section{Introduction}
\label{sec:introduction}
Transit network assignment refers to the process in which an origin-destination (OD) matrix of traveler demand is assigned to a set of transit routes that connect each OD pair. This problem plays a vital role in the modeling and evaluation of transit systems, with rich literature covering both theoretical developments and a wide array of applications. The route (or, more generally, the path) choice of transit users is a key element of transit assignment models. Traditional approaches to route-choice modeling tend to be based on conventional transportation modes, such as private cars or fixed line and schedule public transport services (FIX). To accommodate simulation-based evaluation of novel shared-transport services and demand scenarios, transit route-choice models have been rapidly developing. This can be attributed to an increased attention towards resource-sharing and transit systems as a core component in sustainable urban development, together with innovations within Intelligent Transport Systems (ITS) and real-time information provision to both travelers and operators. Furthermore, new data sources with higher degrees of temporal and spatial richness (e.g., mobile phone, automatic vehicle location [AVL], automatic passenger counting) have become increasingly available for tracking the real-time evolution of both supply and demand. This enables more disaggregate approaches to modeling and forecasting the movements and interactions of individual travelers and vehicles under alternative scenario settings and intervention strategies \citep[]{IbarraRojas2015,Koutsopoulos2019}. 

The term \emph{flexible transit} (FLEX) refers to public transit services characterized by flexible routing and scheduling adapted to traveler needs at a within-day temporal scale. In the literature, FLEX can refer to many service types ranging from door-to-door shared taxi-like services \citep{Fagnant2014,Cebecauer2021} to semi-flexible services with partially fixed routes and timetables that allow for demand-responsive dynamic fleet management \citep{Errico2013}. In a recent survey, \cite{Vansteenwegen2022} compiled a wide range of terms referring to such systems, for example \emph{on-demand}, \emph{dial-a-ride}, \emph{demand-adaptive}, \emph{demand-responsive}, \emph{flex-route}, \emph{flexible}, \emph{hybrid}, or \emph{variable-type} to name a few. 
Generally, FLEX is found to be most beneficial in scenarios when travel demand is low (e.g., in more rural or suburban areas) and/or more variable (e.g., where there are considerable differences between peak and off-peak hours) \citep{Potts2010,Soerensen2021} but are in practice difficult to economically sustain due to difficulties in spreading the cost of individual trips over a greater number of travelers \citep{Ferreira2007,Davison2012}. Interest in FLEX systems has grown in the past decade with the emergence of smartphone-enabled ride-pooling services (e.g., Uber) and technological progress in the development of automated vehicles (AVs). With great opportunities for real-time fleet coordination and prospective lower per-vehicle operational costs \citep{Boesch2018}, many conceptual FLEX use cases for AVs have been proposed in recent years \citep{Narayanan2020}.

Agent-based simulation models have been utilized in evaluating a broad FLEX service design space (both with and without the inclusion of AVs), with varying levels of integration with FIX, as well as different network topologies and demand settings at different spatial and temporal scales \citep{Ronald2015,Markov2021}. In relation to public transit, novel FLEX services evaluated in the literature can be classified as: (i) independent of traditional public transit (e.g., as a shared trip alternative to individual-use taxis or private cars) \citep{Fagnant2014,Martinez2014,Bischoff2016,Liu2017,AlonsoMora2017,Markov2021}, (ii) as a replacement for FIX \citep{Winter2018,Jaeger2018,Berrada2021,narayan2021scalability} or as a competing alternative \citep{Liu2019a,Hoerl2021}, and (iii) as an alternative that may be combined with FIX (e.g., feeder/last-mile) to construct a complete trip from origin to destination \citep{Horn2002,Scheltes2017,Moorthy2017,Salazar2018,Shen2018,Wen2018,Narayan2020,Leffler2021}. While performance measures and motivations for these studies vary, they tend to conclude with similar messages; to achieve a sustainable FLEX service from passenger-oriented, operator-oriented, or societal perspectives, it is crucial that this service is efficiently integrated with existing FIX systems rather than introduced as a directly competing alternative \citep{Narayanan2020, Soerensen2021}. 

The characteristics of FIX or FLEX services, e.g., in terms of expected in-vehicle time, waiting time, crowding, and reliability, may inherently be better suited to the individual preferences of a traveler during different legs (e.g., first-mile or last-mile) of their trip, and in light of new information as their trip progresses. In this sense, it is important to understand how travelers potentially combine FIX and FLEX public transport services in different ways in the evaluation of such systems. Agent-based simulation frameworks utilized for this purpose are developing hand-in-hand with envisioned use cases of AVs and FLEX services. The incorporation of integrated FIX and FLEX route choice models and dynamic assignment techniques within these frameworks are, however, still at their earlier stages of development.

\section{Literature review}
\subsection{Transit route choice and assignment}
Modeling transit route choice involves both representation of the transit system (e.g., in terms of both tactical and operational policies, vehicles types, and transit infrastructure) and how users interact and potentially give rise to congestion (e.g., in the form of in-vehicle crowding, denied boarding, queuing) within the transit system. More recently developed methodology for modeling mode/route choice within studies of FLEX systems largely builds on previous approaches for conventional transit. \cite{Liu2010} categorize route-choice models within the context of transit assignment into three groups: (i) static transit assignment, (ii) within-day dynamic transit assignment, and (iii) emerging approaches. 

The first category, static transit assignment, encompasses methods based on shortest-path estimation (all-or-nothing assignment to the minimal cost route for a given OD pair) and user equilibrium-based assignment. The equilibrium assignment problem is often solved through fixed-point iterations to estimate the travel cost of available routes under capacity constraints, typically using a variant of the method of successive averages (MSA) \citep{Sheffi1982} to update solutions between iterations. Seminal works (see e.g., \cite{Dial1971,Spiess1989}) also find different ways of dealing with the well-known ‘common lines’ problem of transit assignment (travelers at a transit stop often choose opportunistically between several competitive transit lines with potentially different routes) and the task of generating a ‘reasonable’ choice-set of alternative transit routes for each OD (i.e., eliminating paths that are highly unlikely to ever be considered by a traveler) while avoiding the computational intractability of enumerating all possible paths. To incorporate the behavioral aspects of route choice (for example different levels of knowledge, familiarity, or preference of travelers for different routes), these methods are combined with stochastic assignment and discrete random utility maximization (RUM) choice models. 

The adaptive behavior of travelers is intuitively important to model in the presence of FLEX services where, in comparison with FIX, the routing and scheduling of vehicles is even more dependent on traveler requests and supply-demand imbalances at a closer to real-time temporal resolution. Compared to traditional transit assignment models, a FLEX operator can also be viewed as more of a within-day decision-maker that anticipates and responds in real-time to the within-day decision 'events' of travelers. Within-day dynamic transit assignment models relax assumptions regarding the static distributions of travel times and choice behavior of travelers and incorporate the within-day time dimension in route choices \citep{Cats2016}.

Emerging approaches enhance these methods further by attempting to include additional complexities in choice behavior, for example, incorporating the bounded rationality of transit users \citep{Jiang2021}. The inclusion of additional adaptive choice dynamics and the integration of new data sources into route-choice models is a trend that has continued within the past decade \citep{Chen2016}. Such studies are often inspired by research originating in cognitive psychology \citep{Leong2012} and may include how travelers integrate various information sources into their decision making (e.g., smartphone journey-planners, timetables, and both individual and collective experiences) together with models for habit formation and learning \citep{Bogers2007,Cats2020}. 

In simulation-based evaluations of FLEX (whether as an independent mode or as part of a combined FIX and FLEX service), demand may be modeled exogenously, i.e., that demand for the FLEX service is independent of the level-of-service (LoS) provided (e.g., \cite{Shen2018,Jaeger2018}), or endogenously, i.e., that demand for the FLEX service is dependent on LoS provided (e.g., \cite{Atasoy2015,Archetti2017,Berrada2021}). A FLEX system typically does not exist in isolation, but rather as a part of a public transit system as a complementary or competing alternative to other modes, and it is reasonable to model demand for such services endogenously. The vast majority of studies, however, focus on FLEX systems as an independent mode or replacement to conventional modes and assume fixed and inelastic demand for such services \cite{Vansteenwegen2022}.

\subsection{Flexible transit simulation with endogenous demand}
Among studies that model demand for FLEX endogenously, however without integration with FIX, \cite{Atasoy2015} explore the dynamic allocation of a joint fleet of vehicles to individual use taxi, shared-taxi, and mini-bus services. \cite{Archetti2017} model the choice between a door-to-door FLEX shuttle system, FIX, and private car for a wide range of grid-structured road and transit networks and spatio-temporal demand patterns. \cite{Berrada2021} investigate the socioeconomic and economic impact of partially replacing FIX services with a stop-based FLEX service with varying fleet compositions (i.e., number of vehicles and vehicle sizes) based on a real-life case study of such a scenario in Paris. In modeling predicted demand for each FLEX service type, \cite{Atasoy2015} adopt a multinomial logit (MNL) model using projections of fare and perceived utility of travel to estimate traveler choices in response to re-optimized fleet allocations. The choice-set presented to each traveler is determined by a set of constraints based on the capacity of each service and a parameter that defines the departure and arrival time flexibility of travelers. To incorporate individual perceptions of 'best route', \cite{Archetti2017} randomize individual LoS preferences including a FIX user class, however assume travelers have perfect knowledge of all routes and thus assign travelers statically to each service based on shortest travel time. Similar to \cite{Archetti2017}, \cite{Berrada2021} assume perfect knowledge of all routes but calculate demand for FIX and FLEX alternatives using a deterministic function of the ratio of the average generalized travel costs (GTCs) of the two choice alternatives for each OD stop pair. An 'elasticity' parameter is also used as an exponent to the GTC ratio to model the sensitivity to cost differences in resulting mode-split. Both \cite{Atasoy2015} and \cite{Archetti2017} do not consider traffic and transit congestion effects on travel time, and an all-or-nothing assignment is used. \cite{Berrada2021} also assume constant speeds for vehicle and traveler movements. Experienced waiting and in-vehicle times are, however, modeled as dependent on levels of demand for FLEX at each OD, which is updated using a simulation framework including an algorithm for individual FLEX vehicle assignments. The fixed-point assignment equilibrium problem is solved using an MSA for updating the GTCs.

Several recent studies have also been dedicated to modeling FLEX systems with endogenous demand that also model their integration with FIX systems \citep{Wen2018,Pinto2020,Narayan2020}.

\cite{Wen2018} evaluate the feedback loop between service performance and demand for an integrated, on-demand shared automated vehicle (SAV) + FIX service and alternative conventional modes. A nested-logit model (referred to as the 'status-quo' choice model in the paper) is estimated based on a survey of existing modes (walk, bike, car, taxi, and a combined transit nest including bus, rail, and first-mile to rail park/kiss and ride) in a major European city. An SAV to rail service is added as a synthetic mode alternative within the transit nest, with a utility function based on parameters from the estimated model and complementary assumptions and operational cost estimates relating to SAV services. An agent-based simulation platform is used to obtain the LoS of the SAV + rail alternative with static travel times based on predicted demand, which is fed into the nested-logit model iteratively. The equilibrium assignment problem is solved using an MSA update on mode-split output of the choice model between days. Results from a case study centered around a high-frequency commuter rail station suggest that allowing for pre-booked requests, combining fare with transit, and offering an SAV + rail system are effective measures to attract private car trips to more sustainable transit alternatives. The authors highlight the need, however, to investigate combined FIX and SAV + FIX in strategic planning to mitigate poaching demand from other FIX services.

\cite{Pinto2020} provide an extensive study of the trade-offs between allocating resources to FIX service patterns (e.g., short-turning) and frequencies and operating a door-to-door SAV mobility service that can be used both as a first/last-mile service to FIX and as an independent mode. The authors formulated a bi-level joint transit network design and SAV fleet sizing model that incorporates time-dependent mode/route choice in the lower-level problem of mode choice and assignment. A multi-modal (active modes, bus, rail) dynamic transit assignment model \citep{Verbas2015} is used to obtain performance metrics for the FIX alternatives, combined with an SAV simulator \citep{Hyland2018} to obtain performance metrics for SAVs per OD and time period. The performance metrics for FIX, SAV, and SAV + FIX services are fed into an MNL mode-choice model in an iterative equilibrium assignment procedure based on mode and route choices. In a case study of the Greater Chicago metropolitan area for existing FIX demand, the authors find that it is possible to improve the LoS provided by the existing transit network by reallocating resources away from FIX patterns with low utilization rates toward SAV services and high demand FIX patterns.

\cite{Narayan2020} develop a model for integrated FIX and FLEX route choice that allows for a combination of FIX and FLEX modes in a single trip. The model is implemented within the agent-based simulation framework MATSim \citep{Horni2016} and applied to a case study of the city of Amsterdam to study resulting mode-split between private car, walking, biking, FIX, FLEX, and FIX + FLEX modes and sensitivities to FLEX service design variables. To limit the choice-set of integrated public transport routes (combining walking/biking, FIX, and FLEX legs of a trip), a rule-based algorithm is proposed based on parameters for the catchment areas of FIX stops by walking and FLEX modes. Furthermore, multi-modal routes are restricted to a maximum of two transfers, and FLEX trips are either utilized as a direct trip or as a first/last-mile trip to FIX. Demand for each OD is assigned all-or-nothing to the generated route with the highest utility, calculated based on distance-based fare, number of transfers, walking/biking time, in-vehicle time, and waiting time. The default logit-based equilibrium assignment of MATSim is used to allow travelers to update their choice of route, mode, and departure time. Results indicate that LoS improves by offering an integrated FIX and FLEX system. However, users combining FIX and FLEX modes are primarily attracted from existing FIX demand.

In general, there are three key components relevant for the incorporation of mode/route choice into a transit assignment model: (i) the choice-set of alternatives available to each traveler to travel from each origin to a destination, (ii) a decision model of which attributes associated with different alternatives (e.g., travel time, fare, comfort) are considered by a traveler and estimations of how a traveler evaluates these attributes in calculating choice probabilities, and, if an equilibrium approach is utilized, (iii) a day-to-day learning and/or convergence model for how travelers update their evaluations of alternative paths based on the results of previous choices and (potentially) unforeseen deviations from prior anticipations \citep{DiosOrtuzar2011}. The studies described above apply different combinations of these components in modeling demand for FLEX endogenously. Results from case studies of integrated FIX and FLEX services all find that the addition of such a service can improve the LoS provided to passengers without increasing operational costs. However, studies have also shown that the introduction of such services can both encourage or discourage the use of more sustainable modes.

\subsection{Synthesis}
In all studies seen so far where demand for combined FIX and FLEX is considered endogenously, integrated FIX and FLEX systems are combined as a single synthetic mode alternative with predefined paths that may be offered as an alternative to 'FIX only' or non-public transit modes such as private car or taxi services. Route choice decisions are made only once at the origin of the traveler and thus cannot capture en-route adjustments in response to new within-day information that may inform anticipated conditions of downstream paths. In contrast with other approaches, we propose a route choice model where traveler decisions are made in an 'online' fashion, i.e., the path used to reach the traveler's final destination is constructed at different decision nodes corresponding to the within-day actions (walking, boarding, alighting, request-sending) of a traveler. The assignment of travelers to FIX, FLEX, or combined FIX and FLEX services are thus potentially made dynamically within-day. The combinatorial possibilities of a multi-modal trip, including FIX, FLEX, and active-mode legs, can be further extended if needed. Similar to previous work, we generate a static choice-set as an initialization step and apply filtering rules which permit the exclusion of certain route combinations (e.g., transferring from one FLEX leg to another). To model integrated FIX and FLEX demand and supply interactions in all studies mentioned, LoS is used as a feedback measure for travelers to re-evaluate their choices. We adopt a logit-based choice model and MSA approach to include behavioral components in traveler representation and to solve the equilibrium assignment problem under congested conditions.

In this study, we develop a transit route choice model for integrated FIX and FLEX services, i.e., FLEX that can potentially be used as an independent mode from an origin to a destination but also allows travelers to transfer between FIX and FLEX in-vehicle legs of a trip. We assume a single centralized fleet manager of this service that coordinates a fleet of vehicles in response to traveler requests according to a predetermined operational policy, however with no synchronization strategy with the co-existing FIX service. This model is implemented within a dynamic public transit assignment framework representing FIX and FLEX operations. Using the taxonomy of \cite{Vansteenwegen2022}, the FLEX service that we consider in this study is a stop-based, many-to-many system that users access in real-time with no pre-bookings, that is fully-flexible in terms of timetable and semi-flexible in routing. The assignment of FLEX requests to vehicles is done in a dynamic online fashion, meaning that the basic schedule for assigned vehicles can change during operations, even for those already executing a previous assignment.

The contributions of the paper are thus:
 \begin{itemize}
     \item Methodological: A route choice model that allows users to combine walking, FIX, and FLEX services to perform a trip integrated within a day-to-day learning framework.
     \item Substantial: A case study based on a coordinated branched line with bidirectional demand showing how travelers make trade-offs between FLEX options and FIX lines and how these choices evolve with day-to-day learning.
 \end{itemize}
 
The study is divided into two parts. First, the transit simulation framework and extension to model dynamic transit user route choice in the presence of FIX, FLEX, and combined FIX and FLEX mode alternatives is described in Section~\ref{sec:methodology}. In the second part of the paper, a two-link toy network is used to illustrate model properties in a simpler setting in Section~\ref{sec:demo-toy-network}. The combined simulation framework is then applied in a case study based on a real-world coordinated branched transit service in Stockholm characterized by a high-demand trunk section towards the inner-city and lower-demand branches in Section~\ref{sec:application-stockholm-case}. FIX, and FLEX services are run in parallel on branches, serving as a feeder/last-mile service to/from a higher frequency trunk segment. The evolution of experienced LoS and anticipated LoS over time are explored to study the feedback loop between adaptive traveler choices and real-time operational vehicle assignment for different traveler groups: those on the trunk portion, those traveling to or from the branches, and those who travel between the branches. Conclusions and a discussion of limitations and future work is presented in Section~\ref{sec:conclusions}.

\section{Methodology}
\label{sec:methodology}
This section introduces the simulation framework used to model FIX and FLEX operations and extensions to the dynamic transit assignment model to include FLEX alternatives. An overview of the simulation framework is presented in Section~\ref{sec:methodology:simulation-framework}. Descriptions of the path alternative data structure and choice-set generation process that lie at the core of the route choice model are introduced in Section~\ref{sec:methodology:path-definition}. Descriptions of traveler states, actions, and state transitions in the presence of combined FIX + FLEX alternatives are presented in Section~\ref{sec:methodology:traveler-states}. The traveler choice model and how path-sets used in choice probability calculations are associated with different actions are described in Section~\ref{sec:methodology:decision-path-sets} and Section~\ref{sec:methodology:decision-path-sets-fix}. The components of LoS anticipations used in the evaluation of action choices is presented in Section~\ref{sec:methodology:los-anticipation}. Finally, the model for traveler learning between days used in this study is described in Section~\ref{sec:methodology:traveler-information} and Section~\ref{sec:methodology:los-experience}.

\subsection{Simulation framework}
\label{sec:methodology:simulation-framework}
We propose a transit route choice model in the presence of combined FIX and FLEX alternatives. The model is implemented within an agent-based public transit simulation framework, BusMezzo \citep{Toledo2010}, that includes essential components that enable modeling both FIX and FLEX operations and traveler behavior. The modeling framework is event-based and embedded within the mesoscopic traffic simulation model Mezzo \citep{Burghout2004}. 

\begin{figure}[htb]
\includegraphics[width=1\textwidth]{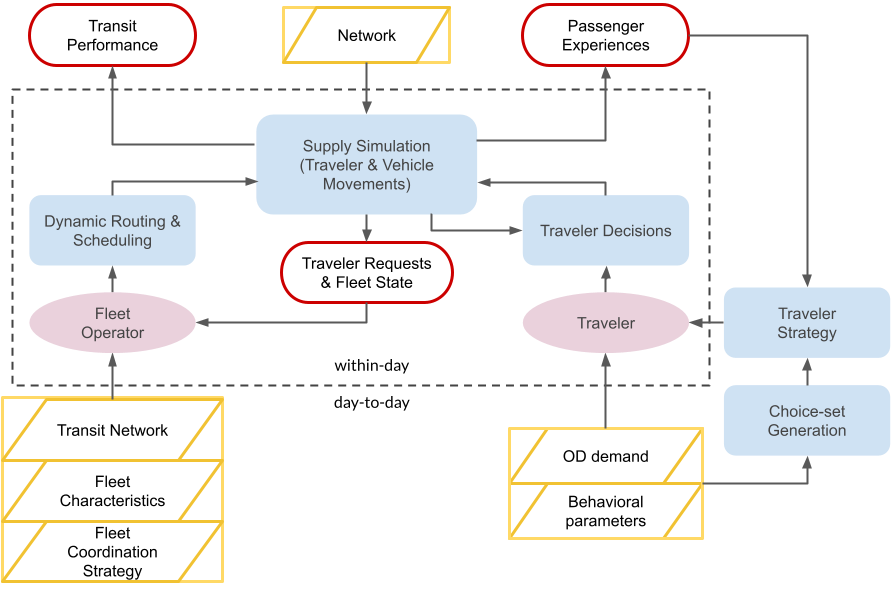}\caption{High-level overview of the simulation framework. Static inputs are marked with yellow boxes, key transit agents with purple ovals, processes with blue rectangles, and final and intermediate process outputs with red stadiums. \label{fig:busmezzo-flowchart}}
\end{figure}

BusMezzo was originally developed to model FIX operations and has been shown to replicate FIX phenomena such as headway variability propagation and bunching \citep{Toledo2010}. In the assessment of FIX operations and dynamic traveler behavior, the model has been applied to evaluate the performance of different real-time holding strategies \citep{Cats2011,Laskaris2018}, short-turning strategies \citep{Leffler2017}, and demand-management strategies with real-time information provision \citep{Drabicki2020,Peftitsi2021}. A high-level overview of the simulation framework is displayed in Figure~\ref{fig:busmezzo-flowchart}.
As shown in Figure~\ref{fig:busmezzo-flowchart}, the simulation framework consists of \emph{within-day} and \emph{day-to-day} processes and feedback loops. Hitherto, the day-to-day learning framework described in \cite{Cats2020} was limited to FIX services. In this study, we introduce modeling functionalities to allow for the representation of FLEX alternatives as part of a dynamic transit assignment model. Static inputs to the framework (marked with yellow boxes) are the underlying road network (e.g., links, nodes, turning-servers), FIX service definitions (e.g., bus lines, stops, schedules, real-time control strategies, fleet characteristics), FLEX service definitions (e.g., possible service routes, stops, vehicle-travel request assignment strategy, empty-vehicle rebalancing strategy, fleet characteristics) and traveler population definitions (e.g., OD demand rates, value-of-time/learning parameters, choice-set filtering parameters). As part of the initialization process, a choice-set generation model is called to generate an initial set of possible path alternatives for each OD, dependent on the available transit network (described further in Section~\ref{sec:methodology:path-definition}). 

Individual traveler agents construct a path from their origin to their destination dynamically within the day with respect to available choice alternatives (as also visualized in Figure~\ref{fig:between-day-decisions}), with intermediate choice probabilities determined by an expected maximum utility. Anticipations regarding the downstream utility of a path alternative are updated based on accumulated experiences learned from day to day. 

Travelers choosing to utilize a FLEX service as part of their chosen path will send a \emph{request} consisting of a desired stop for pick-up, a desired stop for drop-off, and a desired time for departure to the FLEX fleet operator. The FLEX extension to BusMezzo \citep{Leffler2021} allows for switching between different greedy nearest-neighbor assignment heuristics to assign traveler requests made known to a FLEX fleet operator in real-time to vehicles that are coordinated on-demand. Furthermore, alternative empty-vehicle rebalancing strategies may be used. Dependent on the fleet coordination strategy employed by this fleet operator, vehicles are assigned to traveler requests in real-time. The travel times, paths, and choices of individual passengers and network-wide LoS and vehicle utilization measurements are produced as model output for later analysis.

\subsection{Path alternative and choice-set definitions}
\label{sec:methodology:path-definition}
The path alternative data structure lies at the core of the transit route choice model. A path alternative $i$ is defined as an ordered, alternating sequence of walking links $w$ and transit (i.e., FIX or FLEX) links $l$, each connecting pairs of stops $s$, or more generally locations, as follows: 
\begin{equation}
i=\left\{ o,w_{1},s_{2},l_{1},s_{3}\dots s_{2n},l_{n},s_{2n+1},w_{n+1},d\right\}.\label{eq:path-definition}
\end{equation}
The index of each path component denotes its position in this sequence, with the exception of the first and last components, $o$ and $d$, corresponding to the origin and destination of the path, respectively. For example, the stop $o$ denotes the first stop, or origin, of the path alternative $i$, $w_{1}$ denotes the first walking link connecting stops $o$ and $s_{2}$, $l_{1}$ is the first transit link connecting stops $s_{2}$ and $s_{3}$, and so on. A path may include zero-cost walking links that represent the case when a traveler chooses to remain at the current stop rather than walk (e.g., if $o$ and $s_{2}$ in an instance of \eqref{eq:path-definition} refer to the same stop). The number of components in a path alternative is thus fully determined by the number of transit legs $n$ for a given path. 

The global set of path alternatives available to all travelers is generated as part of model initialization. Several filtering rules and logical constraints may be included to limit the number of relevant paths available to each traveler for any OD pair, such as a maximum number of transfers, maximum walking distance, or path dominancy principles. Path components for FIX paths, with the exception of $o$ and $d$, may also be merged into hyperpaths or, more specifically, sets (e.g., $S_{j},W_{j},L_{j}$ for component $j$ of a path alternative) if multiple stops, walking links or transit links that a traveler is considered to be indifferent to, are available for a given segment in a path (described in greater detail in \cite{CatsThesis2011}). For example, suppose several FIX bus lines $l\in L_{j}$ with the same expected travel cost for a given transit link between a pair of stops are available. In that case, a traveler may choose to board any of these lines dependent on whichever bus happens to arrive first (thereby addressing the 'common lines' problem of transit assignment). 

A FLEX leg of a path differs from a FIX leg in the sense that it represents one of many possible routes between stops within the service area of the FLEX operator, which may or may not be realized. In contrast with line-based FIX, a FLEX service operator serves an area containing possible pick-up and drop-off locations rather than a specific route. The FLEX service considered in this study is stop-based, meaning that the set of possible pick-up and drop-off points in a FLEX service area is fixed. In generating the global set of path alternatives, the full set of shortest routes in terms of free-flow travel time between all pairs of stops is generated for the FLEX service area as an initialization step. This set of routes is then used in the generation of traveler choice-sets as for paths including FLEX legs. For simplifying purposes, it is assumed that there are no separately operated FLEX services with overlapping service areas. For this reason, as well as to mitigate the combinatorial explosion of explicit enumeration, in the generation of path-sets, paths with repeating FLEX links are omitted (e.g., if transit link $l_j$ in \eqref{eq:path-definition} is a FLEX link, then $l_{j+1}$ cannot also be a FLEX link). Consequently, FLEX links are always considered distinct from FIX links. This assumption is considered reasonable as several studies have indicated that travelers consider FLEX services as a significantly different mode to FIX \citep{Li2010}. 

\begin{figure}[htb]
\includegraphics[width=1\textwidth]{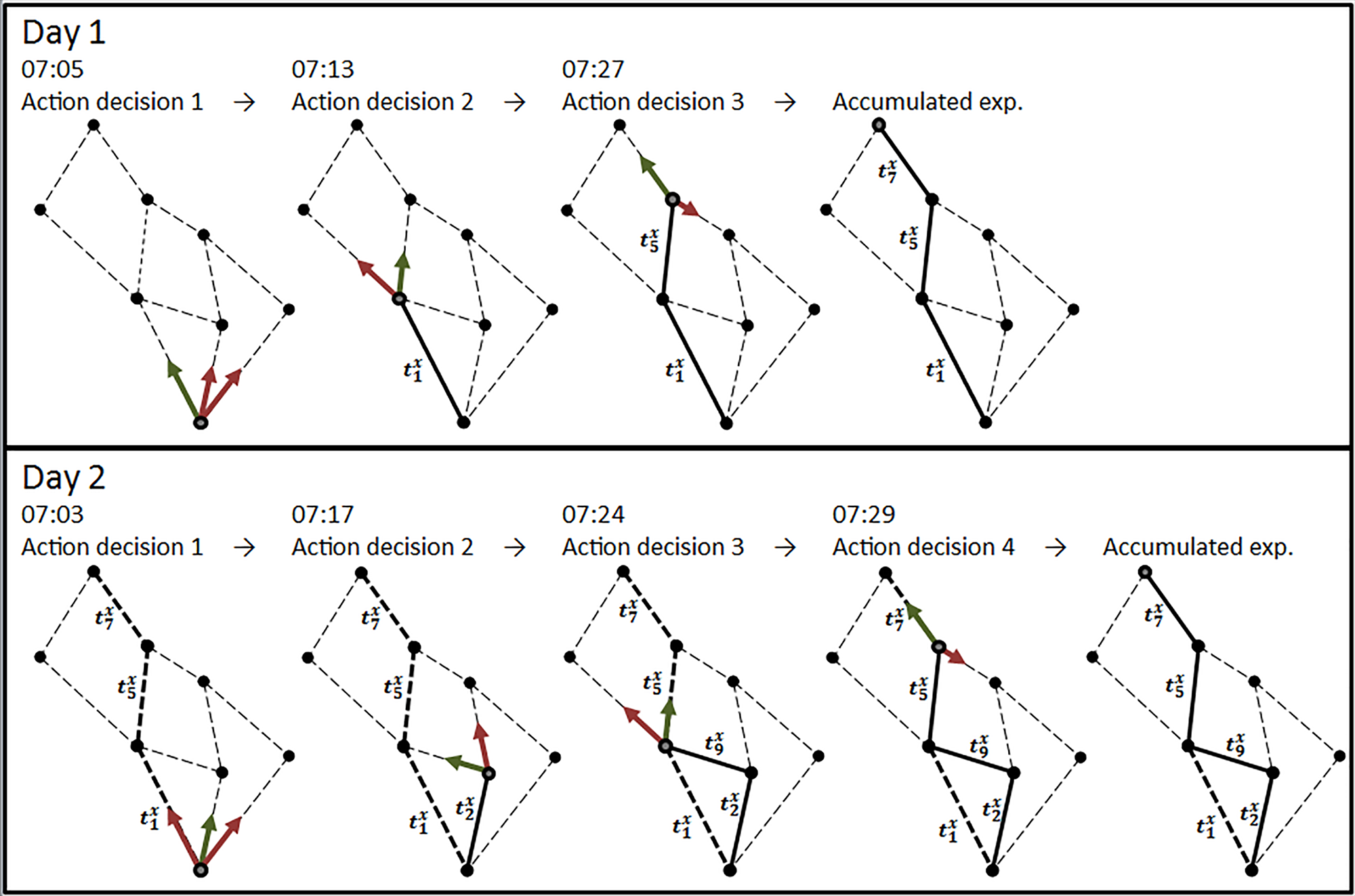}\caption{Example of the sequence of within-day decisions made by a traveler agent and accumulated experiences from the sampled path that inform within-day decisions the following day. Green arrows indicate a chosen action, and red arrows indicate evaluated action alternatives. Source: \citep{Cats2020}\label{fig:between-day-decisions}}
\end{figure}

Let $I^{od}$ denote the set of all path alternatives available to travel between a location $o$ and a destination $d$ utilizing a sequence of walking- and transit links as defined in \eqref{eq:path-definition}. The path that a traveler agent utilizes is not known or selected upfront. Instead, as displayed in Figure~\ref{fig:between-day-decisions}, a traveler makes a sequence of \emph{choices} between alternative walking, boarding, and alighting \emph{actions} at various decision nodes in a trip. Travel experiences (denoted $t^{x}_j$ in Figure~\ref{fig:between-day-decisions}) for each chosen component $j$ of a path are accumulated and inform path choices in the following days (more on this in Section~\ref{sec:methodology:los-experience}).

As displayed in Figure~\ref{fig:between-day-decisions}, at each decision node the set of available path alternatives considered by the traveler will change dependent on their most recently chosen action. The path-set $I_{a}^{od}(t)\subseteq I^{od}$ considered by a traveler when making a choice of action is dependent on the context of the traveler in terms of: (1) the current location $o$, (2) the category of the next action $a$ to be made and (3) the state of the transit network, as well as the traveler's knowledge of this state, at time $t$.

\subsection{Traveler states, actions and state transitions}
\label{sec:methodology:traveler-states}
\begin{figure}[htb]
\includegraphics[width=1\textwidth]{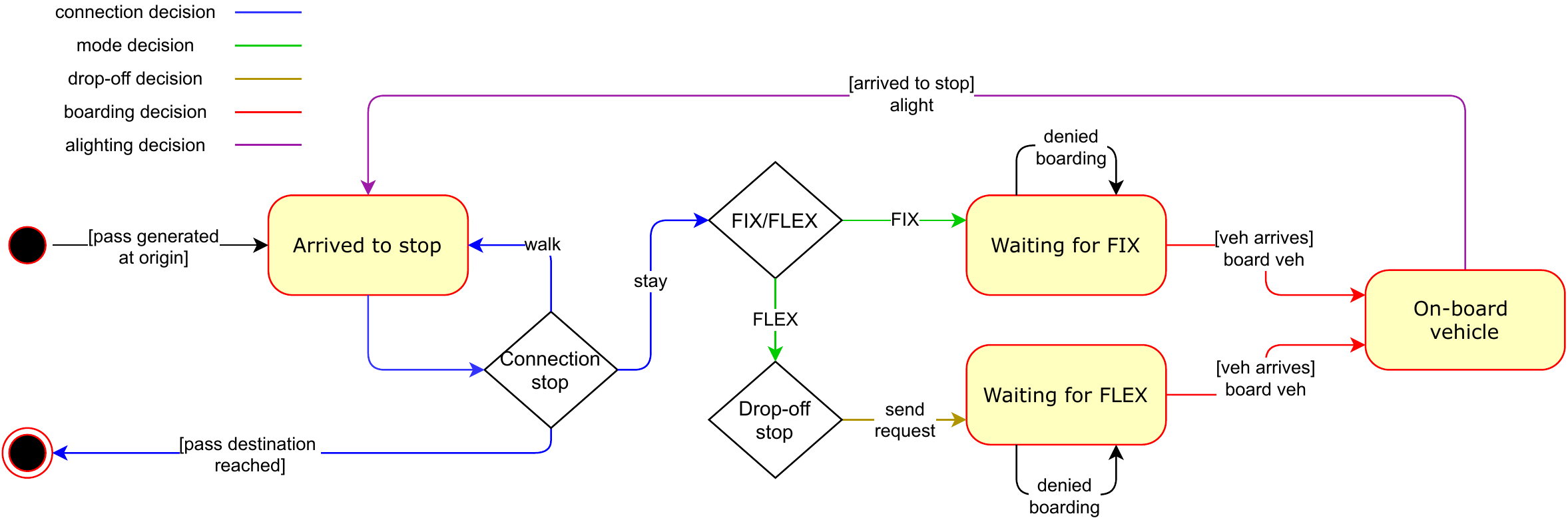}\caption[Traveler states]{Traveler state diagram with FIX and FLEX alternatives available. Traveler states are represented by yellow rounded rectangles, action decision categories with colored arrows, and diamonds where action decisions branch conditional on previous actions. \label{fig:traveler-states}}
\end{figure}

As described in Section~\ref{sec:methodology:path-definition}, a traveler makes a sequence of choices among alternative actions at various decision nodes (or traveler states) to construct a complete trip. A diagram of all traveler states and actions that transition a traveler between these states is displayed in Figure~\ref{fig:traveler-states}.
As shown in Figure~\ref{fig:traveler-states}, there are four states a traveler can be in: (1) having \emph{arrived at a stop} where a decision is made on how to continue their journey to their final destination, (2) \emph{waiting at a stop} where a decision is made to board either a FIX or (3) FLEX vehicle, and (4) sitting or standing \emph{on-board a vehicle} where a decision must be made of which stop to alight at to continue their journey.

Conditional on the current state of the traveler and simulation events (e.g., a vehicle arrives at a stop), five categories of decisions among alternative actions transition the traveler from one state to another. Three action categories connect the two traveler states of having arrived to a stop and waiting for either a FIX or FLEX vehicle to board: (1) a \emph{connection} decision to go to a stop within walking distance (including possibly staying at the current stop) upon arriving to a new location, (2) a \emph{mode} decision of whether to use FIX or FLEX for the next trip leg from the traveler's chosen connection stop and, conditional on FLEX being chosen, (3) a \emph{drop-off} decision of which stop to request a ride to. Note that this drop-off stop can correspond to the traveler's final destination or an intermediate transfer stop. The remaining two action categories follow the state of waiting for a FIX or FLEX vehicle to arrive: (4) a traveler will make a \emph{boarding} decision of whether or not to board an approaching transit vehicle, and, once on-board a vehicle, (5) an \emph{alighting} decision of which stop to alight at. 

Additional assumptions regarding FIX and FLEX service operations and traveler behavior are embedded in the traveler state definitions in Figure~\ref{fig:traveler-states}. It is assumed that a traveler commits to using a FIX or FLEX service once a mode choice has been made (no traveler cancellation). The FLEX service operator also commits to serving the traveler if a trip request is accepted (no operator cancellation). A traveler evaluates a potential trip from a FLEX service operator right after making a connection decision. If FLEX is chosen, the traveler will instantly send a trip request after deciding to walk to a stop to wait for the FLEX service.

\subsection{Connection, mode and drop-off decisions}
\label{sec:methodology:decision-path-sets}
All action decisions described in Section~\ref{sec:methodology:traveler-states} are modeled within the framework of RUM discrete choice models. 
An MNL model is used in calculating the probabilities of action choices. The probability $P_{a,k}$ of traveler $k$ to choose action $a$ from a set of alternative actions $A$ is: 
\begin{align*}
P_{a,k}=\frac{e^{v_{a,k}}}{\sum_{a\in A}e^{v_{a,k}}}
\end{align*}

where $v_{a,k}$ is the expected utility of action $a$ for traveler $k$. Each action $a$ is associated with a downstream path-set, $I_{a}$, that allows the traveler to reach their final destination conditional on this action being chosen. The expected utility $v_{a,k}$ is given by the logsum over the expected utilities of individual path alternatives in $I_{a}$: 
\begin{align}
v_{a,k}=ln\sum_{i\in I_{a}}e^{v_{i,k}}\label{eq:combined-path-utility},
\end{align}
where $v_{i,k}$ is the expected utility of path $i\in I_{a}$ for traveler $k$ (more on this in Section~\ref{sec:methodology:los-anticipation}).

The first choice of action a traveler must make is a connection action, i.e., the choice of which stop to walk to when initiating their trip. This decision is also made each time a traveler alights from a vehicle at a new location. As shown in Figure~\ref{fig:traveler-states}, once a connection decision has been made, it is followed by a mode decision conditional on the chosen connection stop, and a drop-off decision conditional on the choice of mode and connection stop.

In evaluating path alternatives for a connection decision, $o$ corresponds to the current location of the traveler and $d$ the traveler's final destination. Recall the general path definition:
\begin{equation*}
i=\left\{ o,W_{1},S_{2},L_{1},S_{3}\dots S_{2n},L_{n},S_{2n+1},W_{n+1},d\right\}.\label{eq:path-definition-1}
\end{equation*}
A connection decision amounts to choosing between the combined utility of subsets of $I^{od}(t)$, the set of path alternatives available to the traveler to move from current location $o$ to final destination $d$ at time $t$. The path-sets $I_{s^{+}}^{od}\subseteq I^{od}(t)$ associated with each candidate connection stop $s^{+}$ is then given by
\begin{equation*}
I_{s^{+}}^{od}=\left\{ i\mid s^{+}\in S_{i,2}\text{ for some }i\in I^{od}(t)\right\}.    
\end{equation*}
The mode decision (i.e., the choice between FIX and FLEX mode actions) is conditional on the choice of connection stop $s^+$. We can define a function $mode(\cdot)$ that takes a transit link component $L_{i,j}$ of a path $i$ as an argument and outputs whether or not this is operated as a FIX or a FLEX service:
\begin{equation*}
mode(L_{i,j})=\begin{cases}
fix & \text{if a fixed route and schedule exists \ensuremath{\forall l\in L_{i,j}} of path \ensuremath{i} and leg \ensuremath{j}}\\
flex & \text{otherwise}.
\end{cases}
\end{equation*}

The path-sets $I_{s^{+},m}^{od}\subseteq I_{s^{+}}^{od}$ associated with a mode decision are defined conditional on the choice of a stop-mode pair, or action $a=(s^{+},m)$. This action amounts to walking to stop $s^{+}\in S_{i,2}$, using a walking link $w\in W_{i,1},$ with the intention of using mode $m\in\{fix,flex\}$ for the next transit leg $l\in L_{i,1}$ for some path $i$. More formally, the path-sets associated with each mode action are:
\begin{equation*}
I_{s^{+},m}^{od}=\left\{ i\mid m=mode(L_{i,1})\text{ for some }i\in I^{od}_{s^+}\right\}.    
\end{equation*}

If a FLEX link is chosen, a traveler must communicate their intention to use a FLEX service to the operator of this service via a travel request. This request must include a desired time $t^{+}$ and location $s^{+}$ for pick-up, and a desired location for drop-off, $s^{-}$. The pick-up point has already been chosen in the connection decision of the traveler and is associated with a path-set when choosing to utilize FLEX for the next transit leg, i.e., $I_{s^{+},flex}^{od}$. The desired time for pick-up $t^{+}$ is assumed to be as soon as the traveler plans to arrive to stop $s^{+}$, i.e., the current time $t$ plus the expected time it takes to walk from the traveler's current location $o$ to stop $s^{+}$.

\begin{figure}[htb]
    \centering
    \includegraphics[width=1\textwidth]{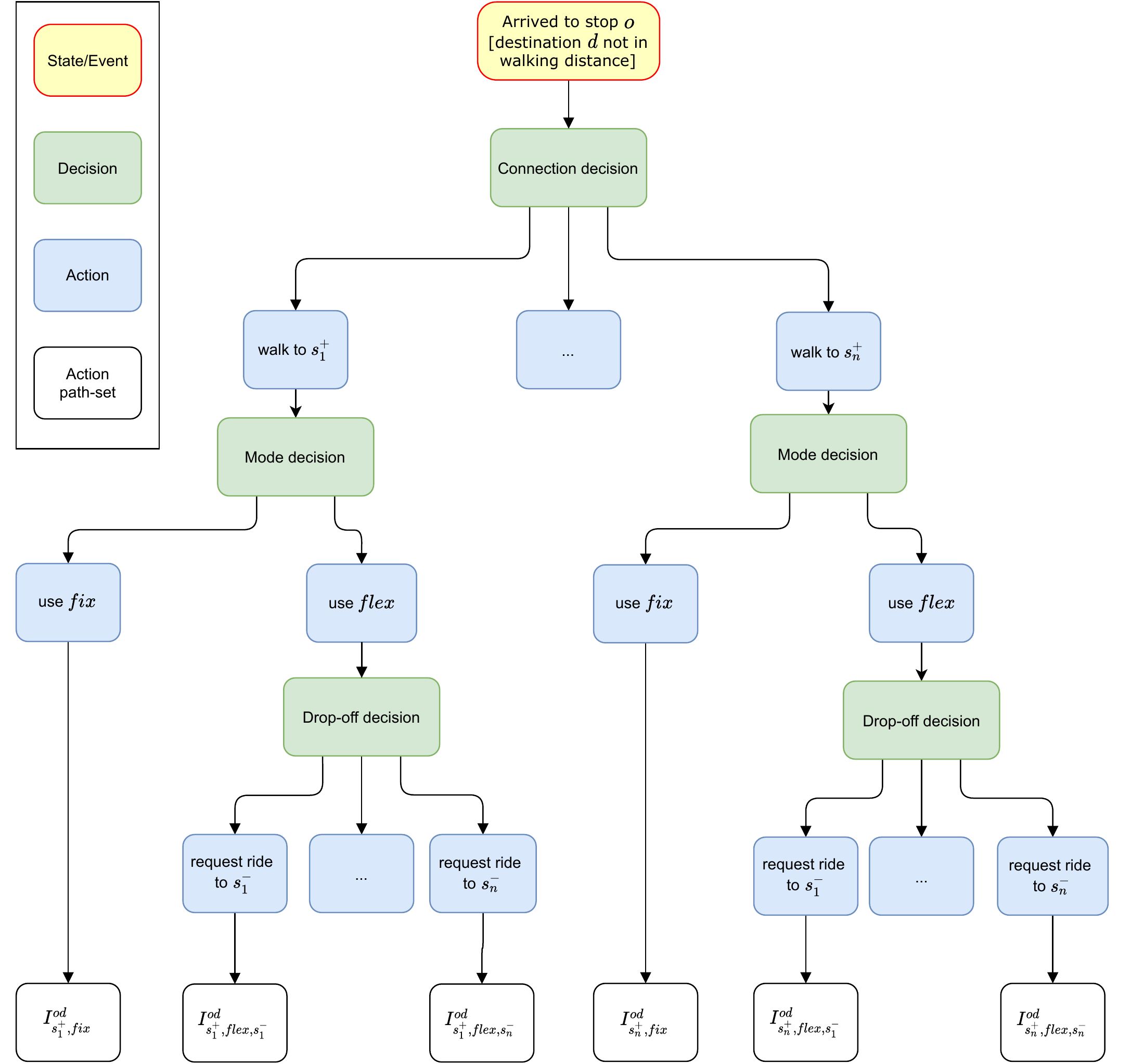}
    \caption{Connection, mode and drop-off decisions.}
    \label{fig:logit-choice-tree}
\end{figure}

If there are several paths in $I_{s^{+},flex}^{od}$, then a decision has to be made by the traveler to choose the most beneficial drop-off point to send a request for. Each drop-off point available after a FLEX transit leg is associated with its own set of path alternatives $I^{od}_{s^{+},flex,s^{-}}\subseteq I_{s^{+},flex}^{od}$. The path-set associated with each candidate drop-off point is given by
\begin{equation*}
I^{od}_{s^{+},flex,s^{-}}=\left\{ i\mid s^{-}\in S_{i,3}\text{ for some }i\in I_{s^{+},flex}^{od}\right\}     
\end{equation*}
The combined utility of each connection, mode and drop-off action and resulting probabilities are then calculated by taking the logsum over the expected utilities of all path alternatives associated with each action (i.e., $I^{od}_{s^{+}},I^{od}_{s^{+},flex}$ and $I^{od}_{s^{+},flex,s^{-}}$) as in \eqref{eq:combined-path-utility}. The choice structure of these decisions can be visualized as a tree as displayed in Figure~\ref{fig:logit-choice-tree}.
Figure~\ref{fig:logit-choice-tree} displays the event triggering a connection decision and the set of alternative actions and associated path-sets for each action.

\subsection{Boarding and alighting decisions}
\label{sec:methodology:decision-path-sets-fix}
Boarding and alighting decisions are invoked if FIX is chosen by a traveler. In this paper, if a decision is made to use FIX, the traveler will walk (if needed) and wait at the chosen stop for the next available FIX service vehicle from a line deemed relevant to board.
The path-set $I_{s^{+},fix}^{od}$, associated with walking to $s^{+}$ to use $fix$, is divided into path alternatives associated with choosing to not board $I^{stay}$ versus choosing to board $I^{board}$ a vehicle arriving from line $l^{arr}(t)$ at time $t$:
\begin{align*}
I^{board}(t) & =\left\{ i\mid l^{arr}(t)\in L_{i,1}\text{ for some }i\in I_{s^{+},fix}^{od}\right\},\\
I^{stay}(t) & =\left\{ i\mid l^{arr}(t)\notin L_{i,1}\text{ for some }i\in I_{s^{+},fix}^{od}\right\} 
\end{align*}
If a traveler has instead decided to use FLEX (i.e., walk to $s^{+}$ to use $flex$ to stop $s^{-}$), then the traveler is committed to boarding any FLEX vehicle that has been assigned to this request. With the assumption that travelers do not cancel a request once sent, there is thus no boarding decision invoked for a traveler in the state of waiting for a FLEX trip.

Similarly, FLEX users are assumed committed to alight at the specific drop-off point for which they sent a request. Again, with the assumption that a traveler will not re-make their decision once a FLEX vehicle has been boarded, no alighting decision is invoked for a traveler in the state of being on-board a FLEX vehicle. Instead, the on-board traveler will alight whenever the chosen drop-off point is reached. For FIX users, the choice of alighting stop is made immediately after a boarding decision has been made. The downstream alternative path-sets for an alighting decision $I^{alight}_{s^-}$ are then determined by the line that was just boarded:
\begin{equation*}
I^{alight}_{s^-}(t)=\left\{ i\mid s^{-}\in S_{i,3}\text{ for some }i\in I^{board}(t)\right\}.  
\end{equation*}

\subsection{Expected utility of a path}
\label{sec:methodology:los-anticipation}
As described in Section~\ref{sec:methodology:decision-path-sets} and Section~\ref{sec:methodology:decision-path-sets-fix}, the probability of choosing an action $a$ is determined by the logsum \eqref{eq:combined-path-utility} over the expected utilities of all path alternatives associated with that action. The expected utility of a path alternative is calculated based on the anticipated GTC associated with traversing stop, walking link and transit link path components. More formally, the expected utility $v_{i,k}$ of path $i\in I_{a}$ for traveler $k$ is given by the weighted sum of anticipated waiting ($\widehat{t}_{j,k}^{wait}$), in-vehicle ($\widehat{t}_{j,k}^{ivt}$), and walking ($\widehat{t}_{j,k}^{walk}$) times over all legs $j\in J_{i}$ of path $i$, and the number of transfers of path $i$, $n_{i}^{trans}$:
\begin{equation}
v_{i,k}=\sum_{j\in J_{i}}\left[\beta_{j,k}^{wait}\widehat{t}_{j,k}^{wait}+\beta_{j,k}^{ivt}\widehat{t}_{j,k}^{ivt}+\beta_{j,k}^{walk}\widehat{t}_{j,k}^{walk}\right]+\beta_{k}^{trans}n_{i}^{trans}, \label{eq:path-utility}
\end{equation}
where $\beta^{wait},\beta^{ivt},\beta^{walk}$ and $\beta^{trans}$ are value-of-time parameters corresponding to each GTC component. The number of transfers for a given path is considered static and known by all travelers. The anticipated walking time associated with walking links is assumed static and uncapacitated, based only on distance and the traveler's average walking speed (realized walking time variability is simulated with stochastic error terms however). The associated cost of waiting- and in-vehicle time at different stages of a trip are dynamic, however, due to confrontations between prevailing demand flows and capacitated supply as well as supply and demand uncertainty (e.g., vehicle occupancy restrictions, on-board crowding, vehicle travel time delays).

Note in the formulation of \eqref{eq:path-utility} that the value-of-time parameters $\beta^{wait},\beta^{ivt},\beta^{walk}$ and $\beta^{trans}$ allows for differentiating between accessing and utilizing transit legs dependent on if it is run by a FIX or FLEX service. In other words, walking or transferring from/to FLEX or FIX transit alternatives, as well as waiting time and in-vehicle time when using FLEX or FIX alternatives, may be perceived differently in the decision-making processes of a traveler.

\subsection{Traveler waiting- and in-vehicle time anticipation}
\label{sec:methodology:traveler-information}
In Section~\ref{sec:methodology:los-anticipation} the anticipated GTC of a path used in the choice model for selecting among alternative actions is described. In this study, the anticipations of waiting- and in-vehicle times (i.e., $\widehat{t}_{j,k}^{wait}$ and $\widehat{t}_{j,k}^{ivt}$ in \eqref{eq:path-utility}, respectively) of traveler $k$ for transit link $j$ of a path are based on two information sources: (1) the accumulated experience ($t_{j,k}^{x}$) of utilizing transit link $j$, and (2) prior knowledge ($t_{j}^{p}$) corresponding to static LoS expectations of transit link $j$.

Prior knowledge anticipations are assumed always known by the traveler before any decision is made (i.e., can inform future decision-making from day 1 and onward). Accumulated experiences are collected within-day, however, only after a decision has already been made. Accumulated experiences are thus only utilized to inform future decisions day-to-day when previously utilized paths are re-evaluated. 

Prior knowledge for FIX transit links is based on the predefined schedule of the service. In the absence of a predefined schedule, an optimistic estimate is used for the anticipated waiting- and in-vehicle times based on prior knowledge for FLEX legs. This could either correspond to a static service guarantee from the operator (e.g., a maximum waiting- and in-vehicle time independent of current fleet and demand status) or a 'best case' evaluation from the traveler's perspective (e.g., immediate service with zero waiting time and the most direct route available with free-flow in-vehicle time). These estimates can be viewed as an exploration parameter used for the first iterations of a traveler's learning process that favors choosing paths that include unexplored FLEX legs. For both FIX and FLEX links, we assume that $t_{j}^{p}$ is only used in the absence of any accumulated waiting- or in-vehicle time experience for a given leg. More precisely, the anticipated waiting- or in-vehicle time $\widehat{t}_{j,k}$ of path component $j$ for traveler $k$ is given by:
\begin{equation}
\widehat{t}_{j,k}=\begin{cases}
t^p_j & \text{if no experience has been gathered by traveler $k$ for leg $j$}\\
t^x_{j,k} & \text{otherwise}.
\end{cases} \label{eq:anticipated_tt}
\end{equation}

\subsection{Accumulated travel time experiences including dynamic congestion effects}
\label{sec:methodology:los-experience}
In Section~\ref{sec:methodology:traveler-information} the prior knowledge and accumulated experience information sources that inform the anticipated GTC calculations are introduced. In this section, we describe the learning model for how accumulated travel time experiences between days is estimated and how additional dynamic congestion effects, such as the discomfort a passenger may experience if denied boarding a full vehicle and the experience of crowding on-board vehicles (dependent on vehicle type and the number of standing/seated passengers) are also considered in the model. 

Learned travel time component anticipations for all decisions in the iterative day-to-day learning process in this study are based directly on the accumulated travel time experiences as defined in \eqref{eq:anticipated_tt}. The accumulated travel time experience of path component $j$ for traveler $k$ at the beginning of day $d$, $t^x_{j,k,d}$ is updated in an MSA formulation. The most recent experiences, $t_{j,k,d-1}$, are discounted as a function of the number of days elapsed (with a discount factor of 1 for the first day). More precisely, the accumulated travel time experiences used in traveler LoS anticipations is given by:
\begin{align}
    t^x_{j,k,d} = t^x_{j,k,d-1} + \frac{1}{d}(t_{j,k,d-1} - t^x_{j,k,d-1}).
    \label{eq:msa_experiences}
\end{align}

To include the additional discomfort induced by denied boarding, a distinction is made between waiting time until the first vehicle the passenger wishes to board ($t_{j,k}^{wait,nominal}$) and additional (unexpected) waiting time ($t_{j,k}^{wait,denied}$) until their next opportunity to board if denied boarding. In registering a waiting time experience, the portion of waiting time due to denied boarding is weighted with an additional penalty $\alpha^{denied}$ for the discomfort experienced. The most recent waiting time experience, $t^{wait}_{j,k,d-1}$, used in the learning update \eqref{eq:msa_experiences} is then given by the weighted sum of these, i.e., $t^{wait}_{j,k,d-1} = t_{j,k,d-1}^{wait,nominal} + \alpha^{denied} \cdot t_{j,k,d-1}^{wait,denied}$.

The total nominal in-vehicle time ($t^{ivt,nominal}_{j,k}$) of passenger $k$ for an in-vehicle leg $j$ of their trip is comprised of the sum of all sub-intervals $h$ of in-vehicle times ($t^{ivt,nominal}_{j,k,h}$) between pairs of stops along the passenger's trip. The additional discomfort experienced when on-board a crowded vehicle is taken into account by multiplying each in-vehicle sub-interval by weights ($\alpha_{h}$) that are a function of crowding level for that segment of the complete trip, i.e., $t^{ivt}_{j,k,d-1} = \sum_h \alpha_{h} \cdot t^{ivt,nominal}_{j,k,h,d-1}$. The crowding level of a given in-vehicle sub-interval may for example be estimated by the ratio of passengers on-board, and the seated and standing capacity of the vehicle.

Since the number of transfers and walking time anticipations of each path are considered static and known beforehand, the learned path component travel time experiences correspond to the average total waiting time (adjusted with additional weight for any waiting time due to denied boarding) and the average crowding-weighted in-vehicle time of different legs of a public transit trip.

\section{Demonstration: Toy network}
\label{sec:demo-toy-network}
\begin{figure}[htb]
\includegraphics[width=0.9\textwidth]{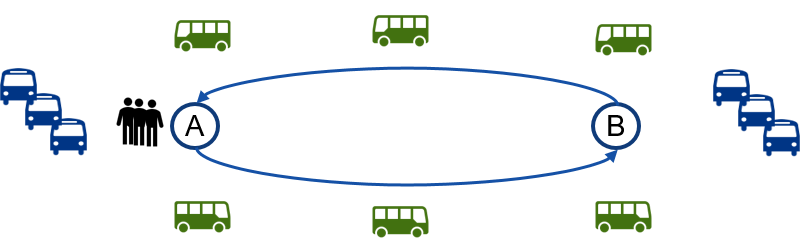}\caption{Two stop and two link network. Two services are available to travelers at stop A traveling to stop B: one fixed headway (10 min) service run with 100 passenger capacity buses (colored green), and one on-demand service run by 10 passenger capacity shuttles (colored blue). Both services use the same links with the same in-vehicle time (30 min) in either direction. \label{fig:2stop-2link}}
\end{figure}

To demonstrate how travelers learn of the trade-offs between waiting time (service headway versus vehicle-travel request assignment) and in-vehicle crowding levels (larger buses versus smaller shuttles) for the FIX and FLEX service types and how this is reflected in resulting mode choices, we construct the following scenario using the setting described in Figure~\ref{fig:2stop-2link}. The physical network consists of two stops (A and B) and two links (AB and BA). Two types of services are available to transport travelers from stop A to stop B: (1) a FIX even-headway (10 min) service run by 100 passenger capacity buses, and (2) a FLEX service with 10 passenger capacity shuttles that assigns vehicle trips to traveler trip requests. FLEX vehicles are assigned to traveler requests every second without rebalancing idling vehicles. Access/egress by walking links is the same for FIX and FLEX services. Both FIX, and FLEX use the same links with the same constant nominal in-vehicle time (30 min) in either direction.

\subsection{Experimental design}
The value-of-time parameters used for calculating the GTC of a path as described in \eqref{eq:path-utility}, as well as in simulation outputs, are generally set in relation to the value of in-vehicle time $\beta^{ivt}$. The weight of perceived waiting time is set to double that of in-vehicle time, $\beta^{wait}=2\cdot\beta^{ivt}$, based on the study of \cite{Wardman2004}. The weight $\alpha^{denied}$ introduced in Section~\ref{sec:methodology:los-experience} to penalize any additional waiting time due to denied boarding is set to 3.5 as detailed in \cite{Cats2016}. 

The in-vehicle crowding factors described in Section~\ref{sec:methodology:los-experience} that are used in this paper are based on the values reported in the meta-study by \cite{Wardman2011}. The load factor is given by the ratio of passengers on-board a vehicle, $q^{onboard}_h$, to the maximum seat capacity of that vehicle, $\gamma^{seats}_h$, for a given time-interval $h$. Based on this load factor an in-vehicle time multiplier is used in weighting passenger in-vehicle time experiences for this time-interval dependent on whether the passenger is seated or standing. Travelers queuing at stops board vehicles in a first-in-first-out manner. The in-vehicle time multiplier for seated passengers, $\alpha^{sit}$, ranges between 0.95 to 1.71 and all passengers that board are assumed to sit if there are unoccupied seats available. The in-vehicle time multiplier for standing passengers, $\alpha^{stand}$, used when all seats are occupied, ranges between 1.78 to 2.69.

To simplify for the sake of demonstration we assume that both FIX and FLEX vehicle types have seats available for each passenger for the toy network scenario. Dwell times ($t^{dwell}$, in seconds) at stops for both FLEX and FIX vehicles are calculated using a linear function of the number of boarding passengers $n^{board}$ and alighting passengers $n^{alight}$ at that stop as estimated in \cite{Dueker2004} for buses without lift operations:
\begin{align}
    t^{dwell} = 5.14 + 3.48n^{board} + 1.7n^{alight}. 
    \label{eq:dwell-time-function}
\end{align}
Besides affecting vehicle running times, $t^{dwell}$ is registered as additional in-vehicle time in traveler experiences.

100 travelers are generated at stop A simultaneously with destination stop B. The arrival time of these travelers is 1 s after the departure time of a FIX bus. Since the capacity of the FIX buses is equal to the total number of travelers, any traveler that chooses FIX will thus experience a waiting time of close to 600 s, the full headway of the service. To simulate varying capacity limitations for the FLEX service, we initialize 1-7 vehicles at stop A before the travelers arrive. If there is capacity at A available for a traveler using FLEX, then the traveler will experience a waiting time of 1 s. If there is no capacity immediately available at A, a vehicle must be sent from stop B to A to serve the traveler. The number of FLEX vehicles initialized at stop B for all scenarios is always sufficient to serve all travelers meaning that, in the worst case, a traveler choosing to use FLEX will experience a waiting time corresponding to the full in-vehicle time it takes for a reactive FLEX vehicle to traverse the link BA, i.e., 30 min.

Mode-choice decisions between FIX and FLEX are mainly dependent on differences in the accumulated experience of waiting time in our demonstration. There are particularly large variations in LoS due to waiting times dependent on how many FLEX vehicles are initialized at stop A and how many passengers choose FLEX on a given day. The influence of anticipated crowding-weighted in-vehicle time in this scenario becomes relevant in conditions when the anticipated waiting time between FIX and FLEX are closer to equal. The system optimum solution for the mode-split over all travelers in terms of resulting total GTC is always that the capacity of the FLEX fleet at A is filled precisely with no overflow with the remaining collection of travelers choosing FIX. In other words, 10\% choosing FLEX is optimal for 1 FLEX vehicle at A, 20\% for 2 vehicles, etc.. 

Finally, with no prior experience of either service, the initial anticipated waiting time for FIX is set to half the headway of the service (i.e., 5 min) and for FLEX an initial exploration parameter of 0 s. The anticipated in-vehicle time for both services is set to the shortest-path nominal in-vehicle time (i.e., 30 min). There is thus an inherent preference towards choosing FLEX over FIX on the first day to stimulate exploration. Each instance is simulated until 75 days have passed with 20 replications per scenario.

\subsection{Results}
\begin{figure}[htb]
    \centering
    \includegraphics[width=1\textwidth]{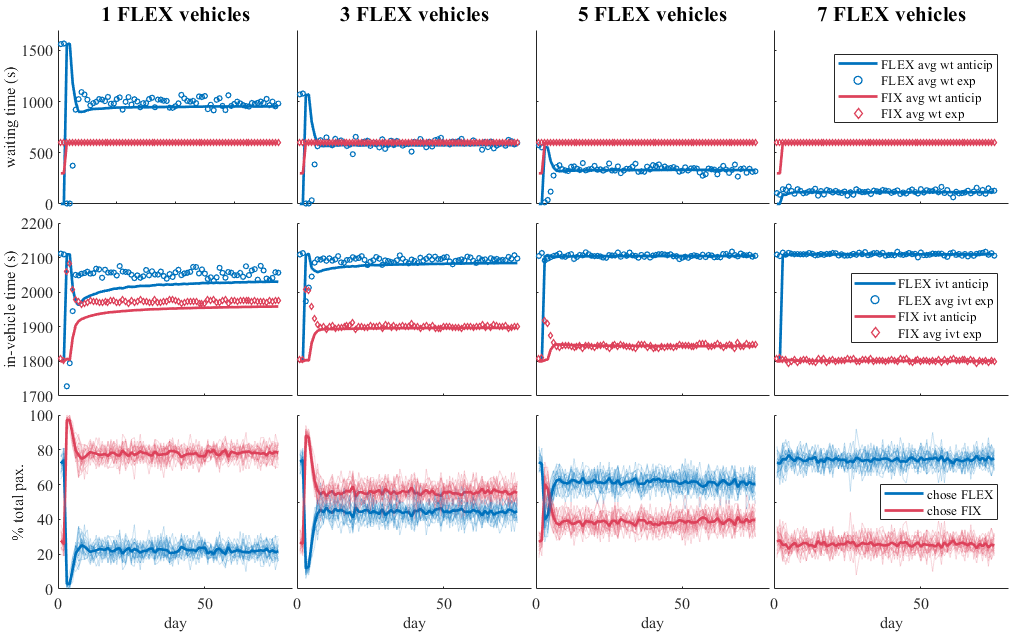}
    \caption{Per-mode (blue for FLEX and red for FIX) anticipated and average experienced waiting time (top row), per-mode anticipated and average experienced (weighted for crowding) in-vehicle time (middle row), and the average share of all travelers that chose FLEX or FIX over days (bottom row). Columns correspond to the size of the FLEX fleet initialized at stop A.}
    \label{fig:toy-learning-modesplit}
\end{figure}

To demonstrate the evolution of average passenger LoS anticipations and long-term convergence behavior, Figure~\ref{fig:toy-learning-modesplit} displays the average (over the 20 replications) learning trajectory of waiting times (top row), crowding-weighted in-vehicle times (middle row) and resulting mode-split (bottom row) for different initial FLEX supply conditions (columns labeled by initial FLEX capacity at stop A).
As can be seen in Figure~\ref{fig:toy-learning-modesplit}, even in this simple setup there is considerable variation and interaction between LoS experiences. The effects of the strong MSA discounting \eqref{eq:msa_experiences} of newer experiences is also apparent, where initial accumulated experiences are retained over a long period of time, and result in smooth learning curves with little oscillation. After the first 7-10 days variations in both mode-split and experiences start to stabilize, which together with quick discounting of newer experiences result in very small changes in LoS anticipations in later days. 

Initial conditions translate into a mode-split of roughly 72\% of travelers choosing FLEX and 28\% choosing FIX on the first day for all FLEX supply variations. For the smallest fleet size of one FLEX vehicle (leftmost column) this means that, on day 1, 10 FLEX passengers experience a waiting time of 0 s, and the remaining 62 must wait 1800 s for 7 vehicles to be sent from stop B. With the exception of the last vehicle to arrive, all of these vehicles are filled to their maximum capacity and the highest crowding factor is used to weight the in-vehicle time of these passengers. FIX passengers always experience a waiting time of 600 s and receive a small discount to their in-vehicle time due to less than half of the seats of the FIX bus being occupied. Since all 28 passengers board the same bus, rather than separate vehicles, a slightly longer dwell time is experienced and registered as in-vehicle time for the FIX passengers compared to FLEX.

The extreme conditions of the first day are followed by a strong increase in anticipated waiting time and perceived in-vehicle time for FLEX and a shift towards FIX the following day (on average 97\% of travelers). The few passengers that now chose FLEX experience a 0 s waiting time and a discounted in-vehicle time, close to the lowest GTC possible for this network. The remaining travelers choosing FIX experience a 600 s waiting time and the highest crowding factor for in-vehicle time. The poor experience of FLEX from the first day is retained over the following days where slowly more passengers, on average 4\% and then 11\%, start to choose FLEX. This is also reflected in a gradually decreasing anticipated waiting time and in-vehicle time which finally converge towards the stabilizing average experiences. In-vehicle times are consistently underestimated with slow convergence to average experiences, however, due to the rapid discounting of new experiences, mode choices converge to a roughly 22\% FLEX and 78\% FIX mode-split. Individual runs oscillate around this within a $\pm6\%$ band due to the random draws of choices.

In the second column of Figure~\ref{fig:toy-learning-modesplit} results for three FLEX vehicles initialized at stop A are displayed. Mode-split and LoS anticipations follow a similar pattern as for the single vehicle case. However, waiting time anticipations converge to an approximately equal level between FIX and FLEX. With similar waiting time experiences, the crowding levels on-board the two vehicle types have a greater impact on choices. The FIX buses are never more than 75\% full, compared to the FLEX shuttles, which are more or less always filled to capacity, resulting in a higher anticipated crowded in-vehicle time for FLEX compared to FIX. Close to equal waiting time anticipation and higher in-vehicle expectations when using FLEX are reflected in the mode split, which converges to roughly 43\% choosing FLEX.

Initial conditions for the 5-7 FLEX vehicle scenarios again result in an average 72\% split in favor of FLEX. This time, however, most FLEX passengers experience a 0 s waiting time instead of 1800 s which results in less extreme oscillation the following days. Waiting time anticipations converge to a lower anticipated waiting time for FLEX compared to FIX. Crowding-weighted in-vehicle times are still consistently higher for the smaller FLEX vehicles when compared to FIX with several vehicles filled to their maximum capacity. However, due to greater differences in waiting times, further weighted at twice the importance in traveler decisions compared to in-vehicle time, the mode-split for the 5 vehicle and 7 vehicle cases converge to close to the capacity of FLEX vehicles at stop A at 60\% and 70\%, respectively.

\section{Application: Stockholm coordinated branched line}
\label{sec:application-stockholm-case}

\begin{figure}[htb]
\includegraphics[width=1\textwidth]{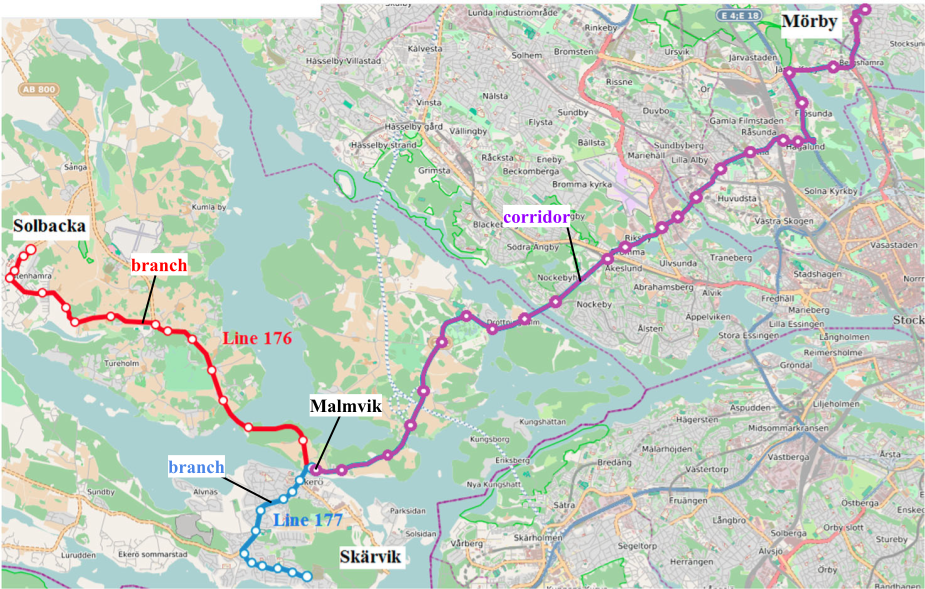}\caption{Lines 176 (red and purple segments) and 177 (blue and purple segment) in Stockholm. The purple segment is common for both lines. The lines merge/split at the stop Malmvik. Figure adapted from \cite{Laskaris2018}, Figure 2.}
\label{fig:drottninghom-network}
\end{figure}

To further investigate model properties at a larger scale and demonstrate how the framework can be applied to evaluate a mixed FIX and FLEX system design, we construct a study based on two existing bus lines in Stockholm: lines 176 and 177. The two lines run with 10 min even-headway during peak hours and form a fork-like trunk-and-branches network that connects rural parts of the Ekerö islands to the more central parts of Stockholm. As displayed in Figure~\ref{fig:drottninghom-network}, the lines run between the branches (colored red and blue) to/from stops Solbacka and Skärvik to the west in the Ekerö region, and form a common corridor (colored purple) to/from Mörby in the northeast. The timetables of the two lines are coordinated such that they run on the corridor section with an even 5 min headway. However, bunching regularly occurs and vehicle utilization on the branch sections is low. This network has been previously studied to evaluate multi-line holding control strategies targeted at reducing bunching \citep{Laskaris2018,Laskaris2021} and shortening the FIX service to the corridor and running a FLEX service on the branches \citep{Leffler2021}. Rather than evaluate the replacement of the FIX services with a FLEX service on branches, as in \cite{Leffler2021}, in this paper we investigate the potential of only partially replacing FIX service on branches with a FLEX service. 
Conceptually, this could provide a more adaptive service on the branches, while improving service regularity on the corridor where demand is higher with more direct connections to the inner-city. This scenario also allows us to further evaluate model properties for a more complex scenario with mixed FIX and FLEX paths.

\subsection{Traveler population definitions}
Table~\ref{tab:demand-input} displays the total demand rates and distribution of demand for the AM (07-10) peak for lines 176 and 177. Given the headway-based operations, the temporal distribution of simulated passenger arrivals in our study is assumed Poisson, i.e., passengers are assumed to arrive at stops independently of anticipations of the timing of vehicle arrivals. The demand rates of these distributions as well as the spatial distribution of arrivals are estimated from aggregated tap-in data obtained from the Stockholm public transport authority (SLL).
\begin{table}[htb]
	\tbl{Total demand rates in average passengers per hour for 3-hour AM peak with directional distribution. The total demand and share of demand per passenger OD categories (corridor-to-corridor [C2C], corridor-to-branch [C2B], branch-to-corridor [B2C], branch-to-branch [B2B]) are also presented for the Eastbound direction toward M\"{o}rby centrum and Westbound direction from M\"{o}rby centrum. Rates and demand shares per passenger OD category are further divided by bus line (i.e., 176, 177).}
    {\begin{tabular}{llccccc} \toprule
     &  &  & \multicolumn{4}{c}{Share of total demand} \\ \cmidrule(l){4-7}
    Line & Direction & \multicolumn{1}{c}{\begin{tabular}[c]{@{}c@{}}Total demand \\      {[}pass./h{]}\end{tabular}} & \multicolumn{1}{c}{C2C} & \multicolumn{1}{c}{C2B} & \multicolumn{1}{c}{B2C} & \multicolumn{1}{c}{B2B} \\
    \midrule
     & Total & 4651 & 48\% & 16\% & 25\% & 12\% \\
    \textbf{Both} & Eastbound & 2485 & 43\% & 0\% & 46\% & 11\% \\
     & Westbound & 2166 & 54\% & 34\% & 0\% & 12\% \\
     \midrule
     & Total & 2477 & 45\% & 15\% & 25\% & 15\% \\
    \textbf{176} & Eastbound & 1369 & 41\% & 0\% & 46\% & 13\% \\
     & Westbound & 1109 & 49\% & 34\% & 0\% & 16\% \\
     \cdashlinelr{1-7}
     & Total & 2174 & 52\% & 16\% & 24\% & 8\% \\
    \textbf{177} & Eastbound & 1117 & 46\% & 0\% & 46\% & 8\% \\
     & Westbound & 1057 & 59\% & 33\% & 0\% & 8\% \\ \bottomrule
    \end{tabular}}
	\label{tab:demand-input}
\end{table}
In describing the demand profile of the two lines in Table~\ref{tab:demand-input}, as well as for later reporting of results, we define the following passenger categories that are spatially grouped based on ODs: corridor-to-corridor (C2C), corridor-to-branch (C2B), branch-to-corridor (B2C) and branch-to-branch (B2B). The directionality of demand for lines 176 and 177 is reasonably balanced. As displayed in Table~\ref{tab:demand-input}, however, in general there is higher demand on line 176 (which is also longer) compared with line 177. For both lines, the majority of demand (around 50\%) in both directions (towards and away from M\"{o}rby) is in the C2C category (i.e., where both the passenger's origin and destination are on the common corridor section of the two lines). The smallest proportion of travelers are within the B2B passenger category, those that have both origin and destination on a branch. The proportion of B2B demand is almost twice as high for line 176 than for line 177.

In a combined FIX and FLEX system there are a multitude of possible path choice compositions from a passenger's origin to their final destination that could be constructed. 
Besides potential computational constraints, many of these compositions are highly improbable to ever be considered an alternative (for example those that include a great deal of backtracking or transfers), yet will still maintain a probability of being chosen and could potentially influence choice behavior and resulting assignment if included in a passenger's choice set. Additionally, it is well known that correlations between route alternatives can induce bias into discrete choice models (e.g., the independence of irrelevant alternatives property of MNL random utility models). 
To focus our analysis on a select group of paths that are characteristic for each of the passenger categories we have defined, we restrict the choice set of each passenger category as displayed in Table~\ref{tab:path-types}.
\begin{table}[htb]
    \tbl{Path type alternatives per passenger category. A one indicates that the path type is available for the corresponding passenger category, and a zero if it is not available.}
    {\begin{tabular}{lcccc} \toprule
     & \multicolumn{4}{c}{Passenger OD category} \\ \cmidrule{2-5}
    Path type & C2C & C2B & B2C & B2B \\ \midrule
    FIX & 1 & 1 & 1 & 1 \\
    FLEX & 0 & 0 & 0 & 1 \\
    FIX-FLEX & 0 & 1 & 0 & 0 \\
    FLEX-FIX & 0 & 0 & 1 & 0 \\ \bottomrule
    \end{tabular}}
	\label{tab:path-types}
\end{table}
As seen in Table~\ref{tab:path-types}, for C2C passengers, only FIX paths without transfers are allowed. Similarly, only direct FIX or FLEX paths are made available to B2B passengers. A direct FIX path is also always available to C2B and B2C passengers if one of the lower frequency lines is chosen. Alternatively, the C2B passengers may transfer from FIX to FLEX and the B2C passengers may transfer from FLEX to FIX. To simplify the choice set further, the first common stop (i.e., Malmvik, as seen in Figure~\ref{fig:drottninghom-network}) between lines 177 and 176 is designated as the only available transfer stop for these passenger categories.

To speed up learning, LoS experiences of alternative paths are shared between travelers that share the same OD. In the learning update described in \eqref{eq:msa_experiences} at the beginning of a day $d$, $t_{j,k,d-1}$ will then correspond to the average experience of a chosen path component $j$ over all travelers $k$ sharing the same OD that chose to use that path component. Similarly, $t^x_{j,k,d-1}$ corresponds to the so-far accumulated experience for travelers sharing the same OD of that path component.

\subsection{Flexible transit operations}
\label{sec:application-stockholm-case:flex-policy}
The FLEX service in this study employs a nearest-neighbor heuristic (described in greater detail in \cite{Leffler2021}) that prioritizes serving stops with the largest cumulative waiting times of passengers when assigning FLEX vehicles to requests and a rule-based rebalancing strategy that aims to maintain an even supply of vehicles at all stops on branches. Note that the aim is not to necessarily optimize the performance of the FLEX service provided but rather represent characteristics of FLEX operations and observe learning output in serving existing public transit users with no pre-bookings. It is common in studies of FLEX assignment algorithms and simulation-based evaluations of FLEX services to include LoS constraints where more costly requests (e.g., low-demand that requires relatively high detouring to serve) may be rejected by the fleet operator to improve assignment efficiency, or to model a maximum acceptable LoS threshold that travelers might tolerate before switching modes or planned activities altogether. To evaluate the alternative system design in our study as a true replacement for a public transit service, however, LoS constraints are not included in determining which requests to accept or reject. Furthermore, once a vehicle is assigned to a FLEX traveler request, the traveler owning this request is also assumed only to board the vehicle assigned to it, i.e., not opportunistically board other vehicles if found available.

When traveler requests are made known to the FLEX fleet operator, they are bundled into trip-plans (consisting of a sequence of stop visits and a route to serve these requests) and assigned to FLEX vehicles dynamically. In this study, if there are no existing trip-plans when a new request is received, a new trip-plan that serves the request directly from origin to destination is generated. The insertion of an additional request into an existing trip-plan is considered feasible if both the stop for pick-up and the stop for drop-off of that request does not require an assigned vehicle to backtrack from already planned stop visits downstream towards the final destination of that trip-plan. Furthermore, a request is only inserted into an existing trip plan if the forecasted load of an assigned vehicle based on already assigned requests does not exceed its maximum passenger capacity.

Assignment calls are made in 'real-time', i.e., every 5 s in this study. For each assignment call, all existing trip-plans are sorted based on the current cumulative waiting time over all requests that have been grouped into each plan. Once sorted, the highest-ranking trip plan (i.e., with the largest cumulative waiting time) is assigned to the nearest on-call vehicle. The next highest-ranking trip-plan is then assigned to the nearest on-call vehicle and so on until there are either no more unassigned trip-plans or no on-call vehicles available.

If on-call vehicles are available after all requests have been assigned to vehicles, then the excess supply may be rebalanced to other stops in anticipation of future demand. In our study, a simple rebalancing strategy is applied. At 10 min intervals throughout the simulation, a rebalancing call is performed. When rebalancing, if available, on-call vehicles are redistributed to all stops (again, the closest on-call vehicle to the stop with the lowest supply) on branches such that an equal supply (i.e., the total number of on-call vehicles at the stop plus the number of vehicles en-route to this stop) at these stops is maintained.

\subsection{Parameter set-up}
In the proposed mixed FIX and FLEX system design, we increase the headway of the two FIX lines on branches from 10 min to 30 min. A third line that only services the corridor with a 7.5 min headway is added, which, in combination with the two original lines, maintains the current combined service frequency of 5 min on the corridor (if headways along the corridor are perfectly synchronized). Reducing the frequency of the FIX lines on the branches means that the FIX bus fleet can be reduced by six buses, each with a maximum capacity of 100 passengers and 44 seats (i.e., an urban articulated bus). For the alternative service design, we exchange the total maximum capacity of these vehicles for 60 smaller FLEX shuttles operating by the policy described in Section~\ref{sec:application-stockholm-case:flex-policy} that only run on the branches, each with a maximum capacity of 10 passengers with 5 seats. 

Access/egress times at stops are considered the same for both FIX and FLEX. For transferring passengers, a FIX cost per transfer $\beta^{trans}$ is set to be roughly equal to 5 minutes of in-vehicle time in line with the study of \cite{Balcombe2004}. To estimate on-board comfort, the in-vehicle crowding factors for standing and seated passengers described in Section~\ref{sec:demo-toy-network} are used.

The running times of both FIX and FLEX vehicles are drawn from log-normal distributions to characterize the variability of vehicle running times, with parameters calculated from AVL data. To model dwell times, we use the same dwell time function \eqref{eq:dwell-time-function} as in the toy network demonstration.

All presented results are averaged over 20 simulation replications per scenario for 100 days of learning. The relative standard errors of the mean for all final day results are then smaller than 2\% for all reported KPIs. Passengers are generated according to the AM peak demand pattern presented in Table~\ref{tab:demand-input} over 3 simulated hours. All results reported are based on choices, travel time experiences, and vehicle utilization results registered within this period. Before the first passenger arrival, a warm-up period is included to distribute FIX vehicles with an even headway along all FIX lines. The FLEX vehicles are initialized as on-call uniformly distributed over all branch stops.

\subsection{Results}
\begin{figure}[htb]
    \centering
    \includegraphics[width=1\textwidth]{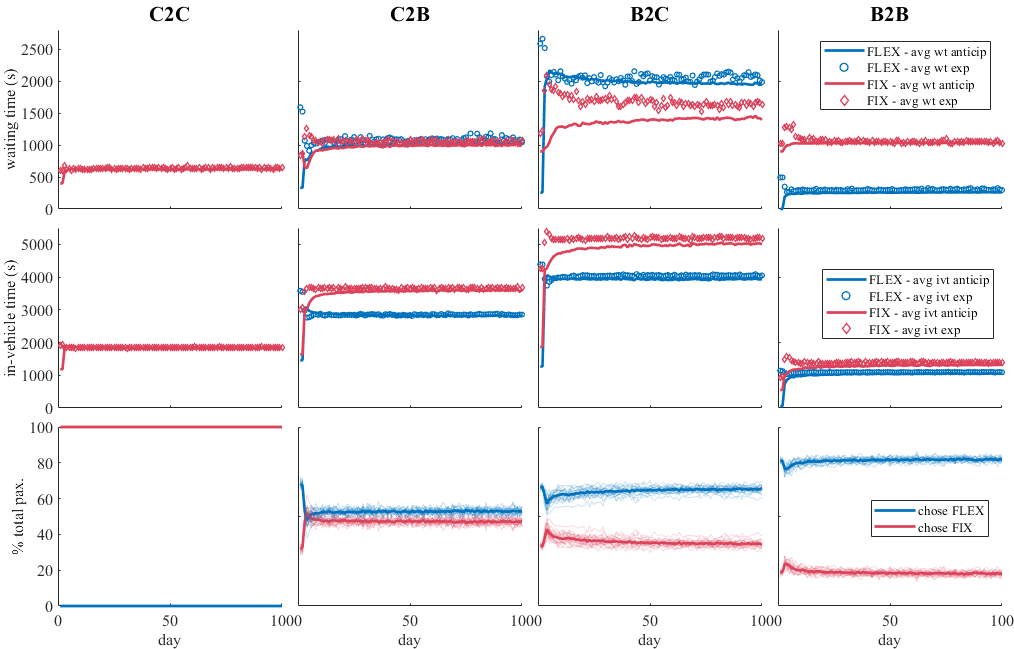} \caption{Learning results from the mixed-mode scenario and AM peak demand. Per path type (FIX- for the C2C category, FIX-FLEX for C2B, FLEX-FIX for B2C, and FLEX- for B2B) waiting time (weighted for denied boarding) anticipation/experience (top row), in-vehicle time (weighted for crowding) anticipation/experience (middle row), and the percentage of all travelers that chose a path that included a FLEX leg or a FIX only path over days (bottom row). Columns correspond to the passenger OD categories.}
    \label{fig:drottningholm-learning-modesplit}
\end{figure}

To explore how evolving LoS expectations translate into mode-choice decisions for the larger case study, learning curves for waiting time including weighted waiting time due to denied boarding, crowding-weighted in-vehicle time, and resulting mode-split in a decision between FIX only paths or paths including a FLEX leg are presented in Figure~\ref{fig:drottningholm-learning-modesplit}. Note that in Figure~\ref{fig:drottningholm-learning-modesplit}, unlike the toy network results in Figure~\ref{fig:toy-learning-modesplit}, the blue 'FLEX' results (as also labeled in the legend) refers to any path that includes a FLEX leg as part of the complete trip, in other words, the FLEX path alternative for each passenger category presented in Table~\ref{tab:path-types}. 
As observed in Figure~\ref{fig:drottningholm-learning-modesplit}, mode-split convergence (bottom row), after initial oscillations, starts to level out after the first 7-10 days with a $\pm6\%$ band due to the random draws of choices. Notably, not all individual mode-split trajectories converge as quickly, however all display a similar long-term trend. The combination of optimistic initial LoS anticipations and rapid discounting of new experiences can be observed in the convergence of the learning curves for waiting time (top row) and in-vehicle time (middle row). This is perhaps most apparent for the B2C passenger learning curves using the FIX-only paths. The initially optimistic anticipations for both path types are retained over successive days. The long-term trend of anticipated experiences moves toward average realized experiences after consistently underestimating them in earlier days, both for waiting times and in-vehicle times. In general, there are higher degrees of variation in waiting time experiences for transferring passengers (in groups C2B and B2C).

In the second row of Figure~\ref{fig:drottningholm-learning-modesplit}, all FIX paths converge to higher weighted in-vehicle times when compared to paths including FLEX for the C2B, B2C, B2B passenger categories, particularly for the transferring passengers. The ratio of passenger-kilometers traveled (PKT) over vehicle-kilometers traveled (VKT) provides a measure of average traveler flow per FIX or FLEX path types. Dividing this by the seat capacity for FIX versus FLEX vehicles indicates crowding levels per path type chosen. The PKT/VKT ratio for FIX trips is 38.74, which, divided by the seat capacity of 44 of FIX buses, would give a load factor of 77\%. For FLEX vehicles, the PKT/VKT ratio is 2.77, which with a seat capacity of 5, yields a load factor of 55\% resulting in higher crowding-weighted in-vehicle times for passengers using FIX.

With perfect even-headway synchronization, the combined frequency of the three lines serving the C2C passengers (leftmost column) would result in an average waiting time of 150 seconds (half the combined headway) for Poisson distributed passengers. This is also set as the initial waiting time anticipation. Similarly, anticipated in-vehicle times are set to the average shortest-path nominal (i.e., uncrowded) in-vehicle times for each OD within this category. However, the actual average waiting time and in-vehicle time experiences are higher, with the anticipation of waiting time converging to roughly 600 seconds. This indicates a high level of bunching of FIX vehicles along the corridor and a higher frequency of denied boarding. Passengers learn to anticipate the consistent deviations from the planned headway and crowding levels.

For the C2B passenger category (second column from the left), travelers have the option of either choosing FIX directly to their final destination (with an even headway of 1800 s) or using any of the FIX lines (with a combined headway of 600 s with perfect synchronization) to the transfer stop Malmvik and take FLEX to their final destination on a branch. The initially optimistic expectations towards FIX-FLEX (300 s waiting time for FIX to the transfer stop + 0 s waiting time for FLEX to the final destination) versus FIX (900 s expected waiting time) result in an initial preference towards the FIX-FLEX path (68\% of all passengers within this category) on the first day. Waiting time anticipations between the two alternative path types converge to close to the same level. Passengers opting to choose FIX experience more crowding for a longer duration of their trip, resulting in higher average in-vehicle times when compared to the FIX-FLEX path. This is reflected in path/mode preferences, which, after initial oscillation, converge slowly together with travel time anticipations to a close to equal number of passengers choosing each path type with a slight preference towards the FIX-FLEX (53\%) path with less crowding but an additional transfer. 

For the B2C passenger category (third column from the left), the choice scenario is between the FIX path from the branches (again with an even headway of 1800 s, however starting on the lower demand branches) and the FLEX-FIX path (again with an initial 0 s expected waiting time for FLEX to the transfer stop Malmvik and 300 s waiting time to transfer to one of the three FIX lines available). Initial anticipations result in a 67\% mode-split in favor of the FLEX-FIX path on the first day. After some initial oscillation in path choices, the combined FLEX-FIX path results in a lower weighted in-vehicle time that compensates for a higher waiting time when compared to FIX only. This results in convergence towards a 66\% preference for the FLEX-FIX path.

Finally, for the B2B passenger category (the rightmost column), the choice is between a FIX direct line with an initial anticipated waiting time of 900 s and a direct FLEX line with an initial anticipated waiting time of 0 s. With a larger difference in initial waiting time expectations, the initial mode-split heavily favors the FLEX paths (81\% of all passengers within this category) on the first day. However, crowding-weighted in-vehicle times converge to similar levels between FIX and FLEX alternatives, with slightly less crowding on-board FLEX vehicles. Average waiting times are consistently far lower for FLEX versus FIX (300 s versus 900 s), and mode-split preferences converge after initial oscillation to 82\% of all passengers choosing FLEX over FIX.

\begin{figure}[htb]
    \centering
    \includegraphics[width=1\textwidth]{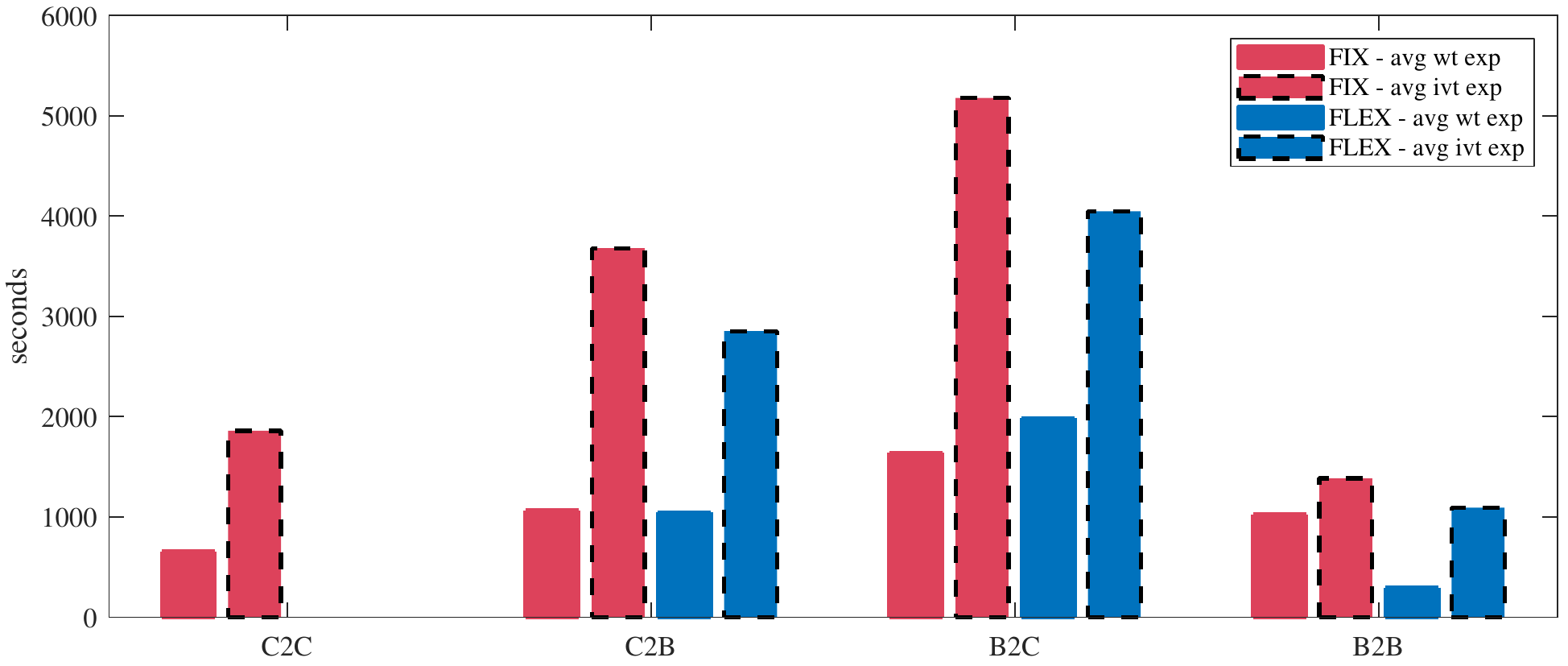} \caption{Final day average waiting time (weighted for denied boarding) and in-vehicle time (weighted for crowding) per passenger OD category (C2C, C2B, B2C, B2B) and per path type (FIX- for the C2C category, FIX-FLEX for C2B, FLEX-FIX for B2C, and FLEX- for B2B).}
    \label{fig:final-day-los}
\end{figure}

The final day waiting times adjusted for denied boarding and the in-vehicle times adjusted for crowding are presented per passenger category in Figure~\ref{fig:final-day-los}. LoS performance is similar between FIX-only paths and paths including FLEX for transferring passengers (C2B, B2C), with a preference towards FLEX mainly due to lower crowding on-board the smaller FLEX shuttles. The B2B passenger category benefits the most from the FLEX service in terms of waiting times compared to FIX-only paths. 

\section{Conclusions and Discussion}
\label{sec:conclusions}
In this paper, a combined FIX (i.e., fixed line and schedule public transit service) and FLEX (i.e., public transit service with flexible routing and scheduling) route choice model is developed and integrated into an existing public transit simulation framework containing core components for modeling FIX and FLEX operations and adaptive traveler behavior. To solve the equilibrium assignment problem and represent traveler day-to-day learning, an MSA approach, where new experiences are discounted by the inverse of the number of days elapsed, is used to guide traveler waiting time and in-vehicle time anticipations towards average realized experiences. A two-link toy network is used to diagnose the implementation of the model and evaluate the evolution of waiting time (including a penalty if denied boarding) and crowding-weighted in-vehicle time trade-offs in mode-split convergence. A case study based on two existing coordinated trunk-and-branches bus lines in Stockholm is then performed to evaluate model properties for a more complex case while demonstrating the usefulness of the combined simulation framework in appraising a broader range of emerging mixed FIX and FLEX service designs and scenarios. 

The two-link toy network intuitively displays a 'worst-case' scenario for travelers utilizing FLEX, with imbalanced and limited supply availability. Large waiting time discrepancies between individual travelers for the capacitated FLEX service are reflected in mode split convergence as the capacity for this service increases. In cases where waiting time anticipations between FIX and FLEX services are similar, differences in crowding-weighted in-vehicle times due to differences in vehicle sizes are also reflected in mode-split, where the larger FIX bus is preferred on average over the smaller FLEX shuttles. The Stockholm case study highlights many important LoS dynamics for first/last-mile mixed FIX and FLEX service design, for example, the uncertainty of waiting times when transferring between congested FIX and FLEX services and potential trade-offs in in-vehicle discomfort due to crowding, dependent on fleet composition. A high degree of bunching and denied boarding was observed for FIX transit users. In particular, for the transferring B2C passengers, high variability in waiting times is observed for travelers transferring between FLEX and FIX and FIX users that have their origin on one of the branches. Both the lack of transfer synchronization to an already congested corridor and the possibility of being denied boarding one of the lower-frequency lines can substantially reduce the LoS experienced by this traveler group.

In the Stockholm case study, it is observed that crowding-weighted in-vehicle times are consistently lower when utilizing multiple smaller shuttles when compared to larger buses. In the absence of FLEX shuttle-specific crowding factors, crowding factors based on the meta-study by \cite{Wardman2011} were applied for both FIX and FLEX passenger in-vehicle experiences. This meta-study was based on experiments for larger vehicles (crowding preferences based on the seated and standing capacity on-board train cars). The crowding effect per passenger or square meter on-board a vehicle with lower passenger capacity is intuitively different. For example, it is questionable if an increase from 3 to 4 seated passengers on-board a ten passenger capacity FLEX shuttle, which would increase the crowding factor for this vehicle from 0.95 to 1.05, is directly comparable to the same increase in crowding level from 33 to 34 passengers for the 100 passenger capacity FIX buses. In future studies, this emphasizes the need for vehicle- and service-type specific value-of-time valuations. The case studies also highlight the need for more explicit consideration of the discontinuous increases in travel cost in LoS anticipations for capacity-restricted FLEX operations. The waiting time distributions of FLEX services can be characterized as being heavily skewed, i.e., travelers tend to either experience the lowest possible travel time for the scenario if FLEX vehicles are available on-call or the longest possible for the network with the learned long-term anticipation converging to the average over these. An interesting line of future work is incorporating waiting time variability more explicitly into the learning processes of the model, for example, through the inclusion of within-day real-time information provision in informing traveler decisions, or more directly in LoS variability valuations (see for instance \cite{AlonsoGonzalez2020, Gerzinic2022}) within traveler GTC functions.

The importance of both initial conditions combined with day-to-day learning dynamics is seen in both the toy network and larger case study. Using the inverse of the number of days elapsed in MSA discounting of new experiences often results in slow convergence that consistently underestimates waiting and in-vehicle times. A dynamic discount factor (for example, the self-regulated averaging method suggested by \cite{Liu2007}) that instead induces more aggressive exploration when there are small differences in LoS anticipations between days and greater discounting when LoS anticipations diverge or taking into consideration additional behavioral factors of learning (such as the distinction between reinforcement-based and belief-based learning in route-choice as suggested by \cite{Bogers2007}) could perhaps be used in future work to improve convergence speed while still pushing anticipations towards an equilibrium.

It is well known that correlations between route alternatives within discrete-choice models can result in inaccurate substitution patterns. Traveler choice-sets were restricted to a predefined set of path types in this study. Although certain correlation structures can be considered within the choice-set generation step, for example, by merging comparable FIX lines into hyperpaths, correlations between routes are not explicitly accounted for. There is a lack of behavioral knowledge on how overlap and similarities among mode-route combinations of fixed and flexible services are perceived. Future research may shed light on the choice-set composition and choice preferences based on empirical behavioral data.

The study presented in this paper leaves several avenues open for future work. The developed simulation framework is expressive in that it allows for experimentation with real-time operational policies and real-time information provided to travelers for individual legs of a complete trip, along with a wide range of behavioral parameters. To simplify the interpretability of results at this stage of model development, traveler adaptation was limited to day-to-day information sources (prior knowledge of each service and accumulated experiences). To further evaluate integrated FIX and FLEX designs for the Stockholm case, experimentation with real-time transfer synchronization strategies \citep{Laskaris2018,Laskaris2021}, combined with real-time information provision to travelers \citep{Cats2020}, is an interesting line of future work. Furthermore, combining the developed simulation framework with day-to-day supply-side re-planning (e.g., trading off fleet size for the FLEX service with the frequency of the individual FIX lines as in \cite{Pinto2020}) could help find a better balance between service types.

\section*{Acknowledgements}
The authors would like to thank Matej Cebecauer for his help and instructive advice in acquiring the demand data used in the second case study and Giorgos Laskaris for sharing the underlying network and fixed transit operations data used in the second case study.

\section*{Funding}
This paper is part of the SMART (Simulation and Modeling of Autonomous Road Transport) project (grant number TRV-2019/27044), financed by the Swedish Transport Administration Trafikverket. The work of the third author was supported also by the CriticalMaaS project (grant number 804469), which is financed by the European Research Council and the Amsterdam Institute for Advanced Metropolitan Solutions.

\bibliographystyle{tfcad}
\bibliography{main}

\begin{thebibliography}{66}
\newcommand{\enquote}[1]{``#1''}
\providecommand{\natexlab}[1]{#1}
\providecommand{\url}[1]{\normalfont{#1}}
\providecommand{\urlprefix}{}

\bibitem[Alonso-Gonz{\'{a}}lez et~al.(2020)]{AlonsoGonzalez2020}
Alonso-Gonz{\'{a}}lez, Mar{\'{\i}}a~J., Niels van Oort, Oded Cats, Sascha
  Hoogendoorn-Lanser, and Serge Hoogendoorn. 2020. ``Value of time and
  reliability for urban pooled on-demand services.'' \emph{Transportation
  Research Part C: Emerging Technologies} 115: 102621.

\bibitem[Alonso-Mora et~al.(2017)]{AlonsoMora2017}
Alonso-Mora, Javier, Samitha Samaranayake, Alex Wallar, Emilio Frazzoli, and
  Daniela Rus. 2017. ``On-demand high-capacity ride-sharing via dynamic
  trip-vehicle assignment.'' \emph{Proceedings of the National Academy of
  Sciences} 114 (3): 462--467.

\bibitem[Archetti, Speranza, and Weyland(2017)]{Archetti2017}
Archetti, Claudia, M.~Grazia Speranza, and Dennis Weyland. 2017. ``A simulation
  study of an on-demand transportation system.'' \emph{International
  Transactions in Operational Research} 25 (4): 1137--1161.

\bibitem[Atasoy et~al.(2015)]{Atasoy2015}
Atasoy, Bilge, Takuro Ikeda, Xiang Song, and Moshe~E. Ben-Akiva. 2015. ``The
  concept and impact analysis of a flexible mobility on demand system.''
  \emph{Transportation Research Part C: Emerging Technologies} 56: 373--392.

\bibitem[Balcombe et~al.(2004)]{Balcombe2004}
Balcombe, Richard, Roger Mackett, Neil Paulley, John Preston, Jeremy Shires,
  Helena Titheridge, Mark Wardman, and Peter White. 2004. \emph{The demand for
  public transport: a practical guide}. Technical {R}eport TRL593. London, UK.

\bibitem[Berrada and Poulh{\`{e}}s(2021)]{Berrada2021}
Berrada, Ja{\^{a}}far, and Alexis Poulh{\`{e}}s. 2021. ``Economic and
  socioeconomic assessment of replacing conventional public transit with demand
  responsive transit services in low-to-medium density areas.''
  \emph{Transportation Research Part A: Policy and Practice} 150: 317--334.

\bibitem[Bischoff and Maciejewski(2016)]{Bischoff2016}
Bischoff, Joschka, and Michal Maciejewski. 2016. ``Autonomous Taxicabs in
  Berlin {\textendash} A Spatiotemporal Analysis of Service Performance.''
  \emph{Transportation Research Procedia} 19: 176--186.

\bibitem[Bogers, Bierlaire, and Hoogendoorn(2007)]{Bogers2007}
Bogers, Enide A.~I., Michel Bierlaire, and Serge~P. Hoogendoorn. 2007.
  ``Modeling Learning in Route Choice.'' Vol. 2014, jan, 1--8. {SAGE}
  Publications.

\bibitem[B{\"o}sch et~al.(2018)]{Boesch2018}
B{\"o}sch, Patrick~M., Felix Becker, Henrik Becker, and Kay~W. Axhausen. 2018.
  ``Cost-based analysis of autonomous mobility services.'' \emph{Transport
  Policy} 64: 76--91.

\bibitem[Burghout(2004)]{Burghout2004}
Burghout, Wilco. 2004. ``Hybrid microscopic-mesoscopic traffic simulation.''
  PhD diss., KTH Royal Institute of Technology.

\bibitem[Cats(2011)]{CatsThesis2011}
Cats, Oded. 2011. ``Dynamic modelling of transit operations and passenger
  decisions.'' PhD diss., KTH Royal Institute of Technology.

\bibitem[Cats et~al.(2011)]{Cats2011}
Cats, Oded, Haris~N. Koutsopoulos, Wilco Burghout, and Tomer Toledo. 2011.
  ``Effect of Real-Time Transit Information on Dynamic Path Choice of
  Passengers.'' \emph{Transportation Research Record: Journal of the
  Transportation Research Board} 2217 (1): 46--54.

\bibitem[Cats and West(2020)]{Cats2020}
Cats, Oded, and Jens West. 2020. ``Learning and Adaptation in Dynamic Transit
  Assignment Models for Congested Networks.'' \emph{Transportation Research
  Record: Journal of the Transportation Research Board} 2674 (1): 113--124.

\bibitem[Cats, West, and Eliasson(2016)]{Cats2016}
Cats, Oded, Jens West, and Jonas Eliasson. 2016. ``A dynamic stochastic model
  for evaluating congestion and crowding effects in transit systems.''
  \emph{Transportation Research Part B: Methodological} 89: 43--57.

\bibitem[Cebecauer et~al.(2021)]{Cebecauer2021}
Cebecauer, Matej, Wilco Burghout, Erik Jenelius, Tatiana Babicheva, and David
  Leffler. 2021. ``Integrating Demand Responsive Services Into Public Transport
  Disruption Management.'' \emph{{IEEE} Open Journal of Intelligent
  Transportation Systems} 2: 24--36.

\bibitem[Chen et~al.(2016)]{Chen2016}
Chen, Cynthia, Jingtao Ma, Yusak Susilo, Yu~Liu, and Menglin Wang. 2016. ``The
  promises of big data and small data for travel behavior (aka human mobility)
  analysis.'' \emph{Transportation Research Part C: Emerging Technologies} 68:
  285--299.

\bibitem[Davison et~al.(2012)]{Davison2012}
Davison, Lisa, Marcus Enoch, Tim Ryley, Mohammed Quddus, and Chao Wang. 2012.
  ``Identifying potential market niches for Demand Responsive Transport.''
  \emph{Research in Transportation Business {\&} Management} 3: 50--61.

\bibitem[de~Dios~Ortúzar and Willumsen(2011)]{DiosOrtuzar2011}
de~Dios~Ortúzar, Juan, and Luis~G. Willumsen. 2011. \emph{Modelling
  Transport}. John Wiley \& Sons.

\bibitem[Dial(1971)]{Dial1971}
Dial, Robert~B. 1971. ``A probabilistic multipath traffic assignment model
  which obviates path enumeration.'' \emph{Transportation Research} 5 (2):
  83--111.

\bibitem[Drabicki et~al.(2020)]{Drabicki2020}
Drabicki, Arkadiusz, Rafa{\l} Kucharski, Oded Cats, and Andrzej Szarata. 2020.
  ``Modelling the effects of real-time crowding information in urban public
  transport systems.'' \emph{Transportmetrica A: Transport Science} 17 (4):
  675--713.

\bibitem[Dueker et~al.(2004)]{Dueker2004}
Dueker, Kenneth~J, Thomas~J Kimpel, James~G Strathman, and Steve Callas. 2004.
  ``Determinants of bus dwell time.'' \emph{Journal of public transportation} 7
  (1): 2.

\bibitem[Errico et~al.(2013)]{Errico2013}
Errico, Fausto, Teodor~Gabriel Crainic, Federico Malucelli, and Maddalena
  Nonato. 2013. ``A survey on planning semi-flexible transit systems:
  Methodological issues and a unifying framework.'' \emph{Transportation
  Research Part C: Emerging Technologies} 36: 324--338.

\bibitem[Fagnant and Kockelman(2014)]{Fagnant2014}
Fagnant, Daniel~J., and Kara~M. Kockelman. 2014. ``The travel and environmental
  implications of shared autonomous vehicles, using agent-based model
  scenarios.'' \emph{Transportation Research Part C: Emerging Technologies} 40:
  1--13.

\bibitem[Ferreira, Charles, and Tether(2007)]{Ferreira2007}
Ferreira, Luis, Phil Charles, and Clara Tether. 2007. ``Evaluating Flexible
  Transport Solutions.'' \emph{Transportation Planning and Technology} 30
  (2-3): 249--269.

\bibitem[Ger{\v{z}}ini{\v{c}} et~al.(2022)]{Gerzinic2022}
Ger{\v{z}}ini{\v{c}}, Nejc, Niels van Oort, Sascha Hoogendoorn-Lanser, Oded
  Cats, and Serge Hoogendoorn. 2022. ``Potential of on-demand services for
  urban travel.'' \emph{Transportation} .

\bibitem[Horn(2002)]{Horn2002}
Horn, M.E.T. 2002. ``Multi-modal and demand-responsive passenger transport
  systems: a modelling framework with embedded control systems.''
  \emph{Transportation Research Part A: Policy and Practice} 36 (2): 167--188.

\bibitem[Horni, Nagel, and Axhausen(2016)]{Horni2016}
Horni, Andreas, Kai Nagel, and Kay Axhausen. 2016. \emph{The Multi-Agent
  Transport Simulation {MATSim}}. London: Ubiquity Press.

\bibitem[Hyland and Mahmassani(2018)]{Hyland2018}
Hyland, Michael~F, and Hani~S Mahmassani. 2018. ``Sharing is caring: Dynamic
  autonomous vehicle fleet operations under demand surges.'' In
  \emph{Transportation Research Board 97th Annual Meeting}, .

\bibitem[Hörl, Becker, and Axhausen(2021)]{Hoerl2021}
Hörl, Sebastian, Felix Becker, and Kay~W. Axhausen. 2021. ``Simulation of
  price, customer behaviour and system impact for a cost-covering automated
  taxi system in Zurich.'' \emph{Transportation Research Part C: Emerging
  Technologies} 123: 102974.

\bibitem[Ibarra-Rojas et~al.(2015)]{IbarraRojas2015}
Ibarra-Rojas, O.J., F.~Delgado, R.~Giesen, and J.C. Mu{\~{n}}oz. 2015.
  ``Planning, operation, and control of bus transport systems: A literature
  review.'' \emph{Transportation Research Part B: Methodological} 77: 38--75.

\bibitem[J{\"a}ger, Brickwedde, and Lienkamp(2018)]{Jaeger2018}
J{\"a}ger, Benedikt, Carsten Brickwedde, and Markus Lienkamp. 2018.
  ``Multi-Agent Simulation of a Demand-Responsive Transit System Operated by
  Autonomous Vehicles.'' \emph{Transportation Research Record: Journal of the
  Transportation Research Board} .

\bibitem[Jiang and Ceder(2021)]{Jiang2021}
Jiang, Yu, and Avishai Ceder. 2021. ``Incorporating personalization and bounded
  rationality into stochastic transit assignment model.'' \emph{Transportation
  Research Part C: Emerging Technologies} 127: 103127.

\bibitem[Koutsopoulos et~al.(2019)]{Koutsopoulos2019}
Koutsopoulos, Haris~N., Zhenliang Ma, Peyman Noursalehi, and Yiwen Zhu. 2019.
  ``Chapter 10 - Transit Data Analytics for Planning, Monitoring, Control, and
  Information.'' In \emph{Mobility Patterns, Big Data and Transport Analytics},
   edited by Constantinos Antoniou, Loukas Dimitriou, and Francisco Pereira,
  229--261. Elsevier.

\bibitem[Laskaris et~al.(2018)]{Laskaris2018}
Laskaris, Georgios, Oded Cats, Erik Jenelius, Marco Rinaldi, and Francesco
  Viti. 2018. ``Multiline holding based control for lines merging to a shared
  transit corridor.'' \emph{Transportmetrica B: Transport Dynamics} 7 (1):
  1062--1095.

\bibitem[Laskaris et~al.(2021)]{Laskaris2021}
Laskaris, Georgios, Oded Cats, Erik Jenelius, Marco Rinaldi, and Francesco
  Viti. 2021. ``A holding control strategy for diverging bus lines.''
  \emph{Transportation Research Part C: Emerging Technologies} 126: 103087.

\bibitem[Leffler et~al.(2021)]{Leffler2021}
Leffler, David, Wilco Burghout, Erik Jenelius, and Oded Cats. 2021.
  ``Simulation of fixed versus on-demand station-based feeder operations.''
  \emph{Transportation Research Part C: Emerging Technologies} 132: 103401.

\bibitem[Leffler et~al.(2017)]{Leffler2017}
Leffler, David, Oded Cats, Erik Jenelius, and Wilco Burghout. 2017. ``Real-time
  short-turning in high frequency bus services based on passenger cost.'' In
  \emph{2017 5th {IEEE} International Conference on Models and Technologies for
  Intelligent Transportation Systems ({MT}-{ITS})}, Jun. {IEEE}.

\bibitem[Leong and Hensher(2012)]{Leong2012}
Leong, Waiyan, and David~Alan Hensher. 2012. ``Embedding Decision Heuristics in
  Discrete Choice Models: A Review.'' \emph{Transport Reviews} 32 (3):
  313--331.

\bibitem[Li and Quadrifoglio(2010)]{Li2010}
Li, Xiugang, and Luca Quadrifoglio. 2010. ``Feeder transit services: Choosing
  between fixed and demand responsive policy.'' \emph{Transportation Research
  Part C: Emerging Technologies} 18 (5): 770--780.

\bibitem[Liu, He, and He(2007)]{Liu2007}
Liu, Henry~X., Xiaozheng He, and Bingsheng He. 2007. ``Method of Successive
  Weighted Averages ({MSWA}) and Self-Regulated Averaging Schemes for Solving
  Stochastic User Equilibrium Problem.'' \emph{Networks and Spatial Economics}
  9 (4): 485--503.

\bibitem[Liu et~al.(2017)]{Liu2017}
Liu, Jun, Kara~M. Kockelman, Patrick~M. Boesch, and Francesco Ciari. 2017.
  ``Tracking a system of shared autonomous vehicles across the Austin, Texas
  network using agent-based simulation.'' \emph{Transportation} 44 (6):
  1261--1278.

\bibitem[Liu et~al.(2019)]{Liu2019a}
Liu, Yang, Prateek Bansal, Ricardo Daziano, and Samitha Samaranayake. 2019. ``A
  framework to integrate mode choice in the design of mobility-on-demand
  systems.'' \emph{Transportation Research Part C: Emerging Technologies} 105:
  648--665.

\bibitem[Liu, Bunker, and Ferreira(2010)]{Liu2010}
Liu, Yulin, Jonathan Bunker, and Luis Ferreira. 2010. ``Transit Users'
  Route-Choice Modelling in Transit Assignment: A Review.'' \emph{Transport
  Reviews} 30 (6): 753--769.

\bibitem[Markov et~al.(2021)]{Markov2021}
Markov, Iliya, Rafael Guglielmetti, Marco Laumanns, Anna
  Fern{\'{a}}ndez-Antol{\'{\i}}n, and Ravin de~Souza. 2021. ``Simulation-based
  design and analysis of on-demand mobility services.'' \emph{Transportation
  Research Part A: Policy and Practice} 149: 170--205.

\bibitem[Martinez, Correia, and Viegas(2014)]{Martinez2014}
Martinez, Luis~M., Gon{\c{c}}alo H.~A. Correia, and Jos{\'{e}}~M. Viegas. 2014.
  ``An agent-based simulation model to assess the impacts of introducing a
  shared-taxi system: an application to {L}isbon ({P}ortugal).'' \emph{Journal
  of Advanced Transportation} 49 (3): 475--495.

\bibitem[Moorthy et~al.(2017)]{Moorthy2017}
Moorthy, Aditi, Robert~De Kleine, Gregory Keoleian, Jeremy Good, and Geoff
  Lewis. 2017. ``Shared Autonomous Vehicles as a Sustainable Solution to the
  Last Mile Problem: A Case Study of {A}nn {A}rbor-{D}etroit Area.''
  \emph{{SAE} International Journal of Passenger Cars - Electronic and
  Electrical Systems} 10 (2): 328--336.

\bibitem[Narayan et~al.(2020)]{Narayan2020}
Narayan, Jishnu, Oded Cats, Niels van Oort, and Serge Hoogendoorn. 2020.
  ``Integrated route choice and assignment model for fixed and flexible public
  transport systems.'' \emph{Transportation Research Part C: Emerging
  Technologies} 115: 102631.

\bibitem[Narayan et~al.(2021)]{narayan2021scalability}
Narayan, Jishnu, Oded Cats, Niels van Oort, and Serge~Paul Hoogendoorn. 2021.
  ``On the scalability of private and pooled on-demand services for urban
  mobility in Amsterdam.'' \emph{Transportation Planning and Technology} 1--17.

\bibitem[Narayanan, Chaniotakis, and Antoniou(2020)]{Narayanan2020}
Narayanan, Santhanakrishnan, Emmanouil Chaniotakis, and Constantinos Antoniou.
  2020. ``Shared autonomous vehicle services: A comprehensive review.''
  \emph{Transportation Research Part C: Emerging Technologies} 111: 255--293.

\bibitem[Peftitsi, Jenelius, and Cats(2021)]{Peftitsi2021}
Peftitsi, Soumela, Erik Jenelius, and Oded Cats. 2021. ``Evaluating crowding in
  individual train cars using a dynamic transit assignment model.''
  \emph{Transportmetrica B: Transport Dynamics} 9 (1): 693--711.

\bibitem[Pinto et~al.(2020)]{Pinto2020}
Pinto, Helen~K.R.F., Michael~F. Hyland, Hani~S. Mahmassani, and I.~Ömer
  Verbas. 2020. ``Joint design of multimodal transit networks and shared
  autonomous mobility fleets.'' \emph{Transportation Research Part C: Emerging
  Technologies} 113: 2--20.

\bibitem[Potts et~al.(2010)]{Potts2010}
Potts, John~F, Maxine~A Marshall, Emmett~C Crockett, and Joel Washington. 2010.
  \emph{A Guide for Planning and Operating Flexible Public Transportation
  Services}. Transportation Research Board.

\bibitem[Ronald, Thompson, and Winter(2015)]{Ronald2015}
Ronald, Nicole, Russell Thompson, and Stephan Winter. 2015. ``Simulating
  Demand-responsive Transportation: A Review of Agent-based Approaches.''
  \emph{Transport Reviews} 35 (4): 404--421.

\bibitem[Salazar et~al.(2018)]{Salazar2018}
Salazar, Mauro, Federico Rossi, Maximilian Schiffer, Christopher~H. Onder, and
  Marco Pavone. 2018. ``On the Interaction between Autonomous
  Mobility-on-Demand and Public Transportation Systems.'' In \emph{2018 21st
  International Conference on Intelligent Transportation Systems ({ITSC})},
  nov. {IEEE}.

\bibitem[Scheltes and de~Almeida~Correia(2017)]{Scheltes2017}
Scheltes, Arthur, and Gon{\c{c}}alo~Homem de~Almeida~Correia. 2017. ``Exploring
  the use of automated vehicles as last mile connection of train trips through
  an agent-based simulation model: An application to {D}elft, {N}etherlands.''
  \emph{International Journal of Transportation Science and Technology} 6 (1):
  28--41.

\bibitem[Sheffi and Powell(1982)]{Sheffi1982}
Sheffi, Yosef, and Warren~B. Powell. 1982. ``An algorithm for the equilibrium
  assignment problem with random link times.'' \emph{Networks} 12 (2):
  191--207.

\bibitem[Shen, Zhang, and Zhao(2018)]{Shen2018}
Shen, Yu, Hongmou Zhang, and Jinhua Zhao. 2018. ``Integrating shared autonomous
  vehicle in public transportation system: A supply-side simulation of the
  first-mile service in Singapore.'' \emph{Transportation Research Part A:
  Policy and Practice} 113: 125--136.

\bibitem[Spiess and Florian(1989)]{Spiess1989}
Spiess, Heinz, and Michael Florian. 1989. ``Optimal strategies: A new
  assignment model for transit networks.'' \emph{Transportation Research Part
  B: Methodological} 23 (2): 83--102.

\bibitem[Sörensen et~al.(2021)]{Soerensen2021}
Sörensen, Leif, Andreas Bossert, Jani-Pekka Jokinen, and Jan Schlüter. 2021.
  ``How much flexibility does rural public transport need? {\textendash}
  Implications from a fully flexible {DRT} system.'' \emph{Transport Policy}
  100: 5--20.

\bibitem[Toledo et~al.(2010)]{Toledo2010}
Toledo, Tomer, Oded Cats, Wilco Burghout, and Haris~N. Koutsopoulos. 2010.
  ``Mesoscopic simulation for transit operations.'' \emph{Transportation
  Research Part C: Emerging Technologies} 18 (6): 896--908.

\bibitem[Vansteenwegen et~al.(2022)]{Vansteenwegen2022}
Vansteenwegen, Pieter, Lissa Melis, Dilay Akta{\c{s}}, Bryan David~Galarza
  Montenegro, F{\'{a}}bio~Sartori Vieira, and Kenneth Sörensen. 2022. ``A
  survey on demand-responsive public bus systems.'' \emph{Transportation
  Research Part C: Emerging Technologies} 137: 103573.

\bibitem[Wardman(2004)]{Wardman2004}
Wardman, Mark. 2004. ``Public transport values of time.'' \emph{Transport
  Policy} 11 (4): 363 -- 377.

\bibitem[Wardman and Whelan(2011)]{Wardman2011}
Wardman, Mark, and Gerard Whelan. 2011. ``Twenty Years of Rail Crowding
  Valuation Studies: Evidence and Lessons from British Experience.''
  \emph{Transport Reviews} 31 (3): 379--398.

\bibitem[Wen et~al.(2018)]{Wen2018}
Wen, Jian, Yu~Xin Chen, Neema Nassir, and Jinhua Zhao. 2018. ``Transit-oriented
  autonomous vehicle operation with integrated demand-supply interaction.''
  \emph{Transportation Research Part C: Emerging Technologies} 97: 216--234.

\bibitem[Winter et~al.(2018)]{Winter2018}
Winter, Konstanze, Oded Cats, Gon{\c{c}}alo Correia, and Bart van Arem. 2018.
  ``Performance analysis and fleet requirements of automated demand-responsive
  transport systems as an urban public transport service.'' \emph{International
  Journal of Transportation Science and Technology} 7 (2): 151--167.

\bibitem[Ömer Verbas and Mahmassani(2015)]{Verbas2015}
Ömer Verbas, {\.{I}}., and Hani~S. Mahmassani. 2015. ``Finding Least Cost
  Hyperpaths in Multimodal Transit Networks.'' \emph{Transportation Research
  Record: Journal of the Transportation Research Board} 2497 (1): 95--105.

\end{thebibliography}

\end{document}